\documentclass{aa}  
\usepackage[varg]{txfonts}
\usepackage{url}
\usepackage{enumitem}

\usepackage{graphicx}
\def\gsim{ \lower .75ex \hbox{$\sim$} \llap{\raise .27ex \hbox{$>$}} }
\def\lsim{ \lower .75ex\hbox{$\sim$} \llap{\raise .27ex \hbox{$<$}} }

\def\beq{\begin{equation}}
\def\eeq{\end{equation}}

\def\sw{{\it Swift}}
\def\fe{{\it Fermi}}
\def\ba{BATSE}

\def\ep{$E_{\rm p}$}

\def\epo{$E_{\rm p,o}$}

\def\liso{$L_{\rm iso}$}
\def\eiso{$E_{\rm iso}$}
\def\ama{$E_{\rm p}-E_{\rm iso}$}
\def\yone{$E_{\rm p}-L_{\rm iso}$}

\def\lf{$\phi(L)$}

\def\psiz{$\Psi(z)$}
\def\sf{$\psi(z)$}

\def\ph{ph cm$^{-2}$ s$^{-1}$}

\begin{document}

\title{Short GRBs at the dawn of the gravitational wave era}
\titlerunning{Short GRBs - Gravitational Waves}
\authorrunning{G. Ghirlanda et al.}
\author{G. Ghirlanda\inst{1}
\thanks{E--mail:giancarlo.ghirlanda@brera.inaf.it}, O. S. Salafia\inst{2,1}, A. Pescalli\inst{4,1}, G. Ghisellini\inst{1},  
R. Salvaterra\inst{5}, E. Chassande--Mottin\inst{6}, \\
M. Colpi\inst{2,3}, F. Nappo\inst{4,1}, P. D'Avanzo\inst{1}, A. Melandri\inst{1}, M. G. Bernardini\inst{7}, M. Branchesi\inst{8,9} \\ 
S. Campana\inst{1}, R. Ciolfi\inst{10,11}, S. Covino\inst{1}, 
D. G\"otz\inst{12}, S. D. Vergani\inst{13,1}, M. Zennaro\inst{14,1}, G. Tagliaferri\inst{1} }
\institute{
$^{1}$INAF -- Osservatorio Astronomico di Brera, via E. Bianchi 46, I-23807 Merate, Italy.\\
$^{2}$Dipartimento di Fisica G. Occhialini, Universit\`a di Milano Bicocca, Piazza della Scienza 3, I-20126 Milano, Italy.\\
$^{3}$Istituto Nazionale di Fisica Nucleare (INFN) -- sede di Milano Bicocca, piazza della Scienza 3, 20123, Milano, Italy. \\
$^{4}$Universit\`a degli Studi dell'Insubria, via Valleggio 11, I-22100 Como, Italy.\\
$^{5}$INAF -- IASF Milano, via E. Bassini 15, I-20133 Milano, Italy. \\
$^{6}$Astroparticule et Cosmologie APC, Universit\'e Paris Diderot, CNRS/IN2P3, CEA/IRFU, Observatoire de Paris, Sorbonne Paris Cit\'e, 75205 Paris, France \\
$^{7}$Laboratoire Univers et Particules de Montpellier, Universit\'e Montpellier 2, 34095 Montpellier, France.\\
$^{8}$Universit\'a degli studi di Urbino "Carlo Bo", Via Saffi 2, 61029 Urbino, Italy\\
$^{9}$INFN, Sezione di Firenze, via G. Sansone 1, 50019 Sesto Fiorentino, Italy\\
$^{10}$Physics Department, University of Trento, Via Sommarive 14, I-38123 Trento, Italy\\
$^{11}$INFN-TIFPA, Trento Institute for Fundamental Physics and Applications, Via Sommarive 14, I-38123 Trento, Italy\\
$^{12}$AIM, UMR 7158 CEA DSM CNRS Universit\'e Paris Diderot, Irfu Service d'Astrophysique, Saclay, France.\\
$^{13}$GEPI, Observatoire de Paris, CNRS, Univ. Paris Diderot, 5 place Jules Janssen, 92190, Meudon, France.\\
$^{14}$Universit\`a degli Studi di Milano, Physics Department, Via Giovanni Celoria, 16, 20133 Milano, Italy. \\ 
}

\date{}


\abstract{
We derive the luminosity function \lf\ and redshift distribution \psiz\ of short Gamma Ray Bursts (SGRBs) using (i) all the available observer--frame constraints (i.e. peak flux, fluence, peak energy and duration distributions) of the large population of \fe\ SGRBs and (ii) the rest--frame properties of a complete sample of SGRBs detected by \sw. We show that a steep $\phi(L)\propto L^{-\alpha}$ with $\alpha\ge2.0$ is excluded if the full set of constraints is considered.  We implement a Monte Carlo Markov Chain method to derive the \lf\ and \psiz\ functions assuming intrinsic \yone\ and \ama\ correlations to hold or, alternatively, that the distributions of intrinsic peak energy, luminosity and duration are independent. To make our results independent from assumptions on the progenitor (NS--NS binary mergers or other channels) and from uncertainties on the star formation history, we assume a parametric form for the redshift distribution of the population of SGRBs. We find that a relatively flat luminosity function with slope $\sim 0.5$ below a characteristic break luminosity $\sim 3 \times10^{52}$ erg s$^{-1}$ and a redshift distribution of SGRBs peaking at $z\sim1.5-2$ satisfy all our constraints. These results hold also if no \yone\ and \ama\ correlations are assumed, and they do not depend on the choice of the minimum luminosity of the SGRB population. We estimate that, within $\sim$200 Mpc (i.e. the design aLIGO range for the detection of gravitational waves produced by NS--NS merger events), there should be 0.007--0.03 SGRBs yr$^{-1}$ detectable as $\gamma$--ray events. Assuming current estimates of NS--NS merger rates and that all NS--NS mergers lead to a SGRB event, we derive a conservative estimate of the average opening angle of SGRBs  $\langle \theta_{\rm jet}\rangle\sim3^\circ$--$6^\circ$. The luminosity function implies a prompt emission average luminosity $\left\langle L \right\rangle \sim 1.5 \times 10^{52}\,\rm{erg\,s^{-1}}$, higher by nearly two orders of magnitude compared to previous findings in the literature, which greatly enhances the chance of observing SGRB ``orphan'' afterglows. Efforts should go in the direction of finding and identifying such orphan afterglows as counterparts of GW events.}

\maketitle

\section{Introduction}
The population of short Gamma Ray Bursts (SGRBs) is still poorly understood due to the relatively few events with measured redshift \citep[see e.g.][for recent reviews]{2014ARA&A..52...43B,2015JHEAp...7...73D}. Available information is rather sparse, but the low density of the close circumburst medium \citep{2013ApJ...776...18F,Fong:2015fp}, the variety of galaxy morphologies \citep[e.g.][]{2015JHEAp...7...73D}, the lack of any associated supernova in the nearby SGRBs and the possible recent detection of a ``kilonova'' \citep{1989Natur.340..126E,1998ApJ...507L..59L,Yang:2015lr,Yang:2015hl,Jin:2016rr,Jin:2015cr} signature \citep{2013ApJ...774L..23B,2013Natur.500..547T}, all hint to an origin from the merger of two compact objects (e.g. double neutron stars) rather than from a single massive star collapse.

However, the prompt $\gamma$--ray emission properties of SGRBs \citep{2009A&A...496..585G,2015JHEAp...7...81G} and the sustained long lasting X--ray emission (despite not ubiquitous in short GRBs - \citealt{Sakamoto:2009lr}) 
and flaring activity suggest that the central engine and radiation mechanisms are similar to long GRBs. Despite still based on a couple of breaks in the optical light curves, it seems that also SGRBs have a jet: current measures of $\theta_{\rm jet}$ are between 3$^\circ$ and 15$^\circ$  while lower limits seem to suggest a wider distribution \citep[e.g.][]{2014ARA&A..52...43B,2015ApJ...815..102F}. Recently, it has been argued that the customary dividing line at $T_{90}=2\,\rm{s}$ between short and long GRBs provides a correct classification for \textit{Fermi} and \textit{CGRO} GRBs, but it is somewhat long for \textit{Swift} bursts \citep{2013ApJ...764..179B}. 

A renewed interest in the population of SGRBs is following the recent opening of the gravitational wave (GW) ``window'' by the LIGO--Virgo discovery of GW150914 \citep{Abbott:2016lr} and by the most recent announcement of another event, GW151226, detected within the first data acquisition run \citep{2016arXiv160604856T,Abbott:2016lr}.  Despite no electromagnetic (EM) counterpart was identified within the large localisation region of these event, there are encouraging prospects for forthcoming GW discoveries to have an EM--GW association, thanks to the aLIGO--Virgo synergy and world wide efforts for ground and space based follow up observations. 

If the progenitors are compact object binary (NS--NS or NS--BH - \citealt[e.g.][]{Giacomazzo:2013fk}) mergers, SGRBs are one of the most promising electromagnetic counterparts of GW events detectable by the advanced interferometers.  
Other EM counterparts are expected in the optical \citep{Metzger:2012fj}, X-ray \citep{Siegel:2016qy, Siegel:2016lq} and radio bands \citep{Hotokezaka:2016uq}.
The rate of association of GW events with SGRBs is mainly determined by the rate of SGRBs within the relatively small horizon set by the sensitivity of the updated interferometers aLIGO and Advanced Virgo \citep{2016LRR....19....1A}. However, current estimates of local SGRB rates range from 0.1--0.6 Gpc$^{-3}$ yr$^{-1}$ (e.g. Guetta \& Piran 2005; 2006) to 1--10 Gpc$^{-3}$ yr$^{-1}$ \citep[WP15]{2006A&A...453..823G,2009A&A...498..329G,2012MNRAS.425.2668C,2014MNRAS.437..649S} to even larger values like 40-240 Gpc$^{-3}$ yr$^{-1}$ \citep{2006ApJ...650..281N,2006A&A...453..823G}\footnote{All these rates are not corrected for the collimation angle, i.e. they represent the fraction of bursts whose jets are pointed towards the Earth, which can be detected as $\gamma$--ray prompt GRBs.}. 

Such rate estimates
mainly depend on the luminosity function \lf\ and redshift distribution \psiz\ of SGRBs. These functions are usually derived by fitting the peak flux distribution of SGRBs detected by BATSE \citep{2005A&A...435..421G,2006A&A...453..823G,2006ApJ...650..281N,2006ApJ...643L..91H,2008MNRAS.388L...6S}. Due to the degeneracy in the parameter space (when both \lf\ and \psiz\ are parametric functions), the redshift distribution was compared with that of the few SGRBs with measured $z$. 
The luminosity function \lf\ has been typically modelled as a single or broken power law, and in most cases it was found to be similar to that of long GRBs \citep[i.e.\ proportional to $L^{-1}$ and $L^{-2}$ below and above a characteristic break $\sim 10^{51-52}$ erg s$^{-1}$ -][D14 hereafter]{2006A&A...453..823G,2008MNRAS.388L...6S,2011ApJ...727..109V,2014MNRAS.442.2342D} or even steeper \citep[$L^{-2}$ and $L^{-3}$ -][WP15 hereafter]{2015MNRAS.448.3026W}. Aside from the mainstream, \cite{2015MNRAS.451..126S} modelled all the distributions with lognormal functions.

The redshift distribution \psiz\ (the number of SGRBs per comoving unit volume and time at redshift $z$) has been always assumed to follow the cosmic star formation rate with a delay which is due to the time necessary for the progenitor binary system to merge. With this assumption, various authors derived the delay time $\tau$ distribution, which could be a single power law $P(\tau)\propto\tau^{-\delta}$ (e.g. with $\delta=1-2$, \citeauthor{2005A&A...435..421G} \citeyear{2005A&A...435..421G}, \citeyear{2006A&A...453..823G}; D14; WP15) with a minimum delay time $\tau_{\rm min}=10-20$ Myr, or a peaked (lognormal) distribution with a considerably large delay (e.g. 2--4 Gyr, \citeauthor{2005AAS...20715803N} \citeyear{2005AAS...20715803N}; WP15). Alternatively, the population could be described by a combination of prompt mergers (small delays) and large delays \citep{2011ApJ...727..109V} or to the combination of two progenitor channels, i.e. binaries formed in the field or dynamically within globular clusters \citep[e.g.][]{2008MNRAS.388L...6S}.

Many past works, until the most recent, feature a common approach: parametric forms are assumed for the compact binary merger delay time distribution and for the SGRB luminosity function; free parameters of such functions are then constrained through (1) the small sample of SGRBs with measured redshifts and luminosities and (2) the distribution of the $\gamma$--ray peak fluxes of SGRBs detected by past and/or present GRB detectors. A number of other observer frame properties, though, are available: fluence distribution, duration distribution, observer frame peak energy. The latter have been considered in \cite{2015MNRAS.451..126S} which, however, lacks a comparison with rest--frame properties of SGRBs as done in this article. Another issue was the comparison of the model predictions with small and incomplete samples of SGRBs with measured $z$. Indeed, only recently D14 worked with a flux--limited complete sample of SGRBs detected by \textit{Swift}. 

The aim of this paper is to determine the redshift distribution \psiz\ and the luminosity function \lf\ of the population of SGRBs, using all the available observational constraints of the large population of bursts detected by the \fe--Gamma Burst Monitor (GBM) instrument. These constraints are: (1) the peak flux, (2) the fluence, (3) the observer frame duration and (4) the observer frame peak energy distributions. In addition we also consider as constraints (5) the redshift distribution, (6) the isotropic  energy and (7) the isotropic luminosity of a complete sample of SGRBs detected by \sw\ (D14). This is the first work aimed at deriving \lf\ and \psiz\ of SGRBs which considers constraints 2--4 and 6--7. Moreover, we do not assume any delay time distribution for SGRBs but derive directly, for the first time, their redshift distribution by assuming a parametric form.

In \S2 we describe our sample of SGRBs without measured redshifts detected by \fe/GBM, which provides observer--frame constraints 1--4, and the (smaller) complete sample of \sw\ SGRBs of D14, which provides rest--frame constraints 5--7. One of the main results of this paper is that the \lf\ of SGRBs is flatter than claimed before in the literature: by extending standard analytic tools present in the literature, we show (\S3) that a steep \lf\ is excluded when all the available constraints (1--7) are considered. We then employ a Monte Carlo code (\S4) to derive the parameters describing the \lf\ and \psiz\ of SGRBs. In \S5 and \S6 the results on the \lf\ and \psiz\ of SGRBs are presented and discussed, respectively, and in \S7 we compute the local rate of SGRBs, discussing our results in the context of the dawning GW era. We assume standard flat $\Lambda$CDM cosmology with $H_0 = 70$ km s$^{-1}$ Mpc$^{-1}$ and $\Omega_{\rm{m}} = 0.3$ throughout the paper. 

\section{Sample selection}\label{sec:sample_selection}

As stated in the preceding section, the luminosity function and redshift distribution of SGRBs have been derived by many authors, by taking into account the following two constraints:
\begin{enumerate}
\item the peak flux distribution of large samples of SGRBs detected by \textit{CGRO}/BATSE or \textit{Fermi}/GBM; 
\item the redshift distribution of the SGRBs with measured $z$. 
\end{enumerate}
\noindent However, a considerable amount of additional information on the prompt $\gamma$--ray emission of SGRBs can be extracted from the BATSE and GBM samples. In particular, we can learn more about these sources by considering the distributions of:
\begin{enumerate}[resume]
\item the peak energy \epo\ of the observed $\nu F_{\nu}$ spectrum; 
\item the fluence $F$; 
\item the duration $T_{90}$.
\end{enumerate}
\noindent Moreover, for the handful of events with known redshift $z$, we have also access to the\footnote{For the sake of neatness, throughout this work we will sometimes drop the ``$\rm{iso}$'' subscript, so that $L_{\rm{iso}}$ and $E_{\rm{iso}}$ will be equivalently written as $L$ and $E$ respectively. For the same reason, the peak energy $E_{\rm peak,obs}$ ($E_{\rm peak,rest}$) of the $\nu F(\nu)$ spectrum in the observer frame (in the local cosmological rest frame) will be sometimes written as $E_{\rm p,o}$ ($E_{\rm p}$).}
\begin{enumerate}[resume]
\item isotropic luminosity \liso;
\item isotropic energy \eiso.
\end{enumerate}

\subsection{Observer-frame constraints: \textit{Fermi}/GBM sample}

For the distributions of the observer frame prompt emission properties (constraints 1, 3, 4, 5) we consider the sample of 1767 GRBs detected by \textit{Fermi}/GBM (from 080714 to 160118) as reported in the on--line spectral catalogue\footnote{\url{https://heasarc.gsfc.nasa.gov/W3Browse/fermi/fermigbrst.html}}. It contains most of the GRBs published in the second (first 4 years) spectral catalogue of \fe/GBM bursts \citep{2014ApJS..211...12G}, plus events detected by the satellite in the last two years. 295 bursts in the sample are SGRBs (i.e.\ with $T_{90}\le$ 2 s).  According to \cite{2013ApJ...764..179B}, for both the \fe\ and \textit{CGRO} GRB population, this duration threshold should limit the contamination from collapsar-GRBs to less than 10\% (see also WP15). 

We select only bursts with a peak flux (computed on 64ms timescale in the 10-1000 keV energy range) larger than 5 \ph\, in order to work with a well defined sample, less affected by the possible incompleteness close to the minimum detector flux. With this selection, our sample reduces to 211 SGRBs, detected by \textit{Fermi}/GBM in 7.5 years within its field of view of $\sim$70\% of the sky.\\
We consider the following prompt emission properties of the bursts in the sample, to be used as constraints of our population synthesis model: 
\begin{itemize}
\item the distribution of the 64ms peak flux $P_{64}$ (integrated in the 10-1000 keV energy range). This is shown by black symbols in the top left panel of Fig.~\ref{fg1};
\item the distribution of the observed peak energy of the prompt emission spectrum \epo\ (black symbols, bottom left panel in Fig.~\ref{fg1});
\item the distribution of the fluence $F$ (integrated in the 10--1000 keV energy range) (black symbols, bottom middle panel in Fig.~\ref{fg1});
\item the distribution of the duration $T_{90}$ of the prompt emission (black symbols, bottom right panel in Fig.~\ref{fg1});
\end{itemize}

Short GRB spectra have a typical observer frame peak energy \epo\ distribution \citep[e.g.][]{2009A&A...496..585G,2011A&A...530A..21N,2014ApJS..211...12G} centred at relatively large values ($\sim 0.5-1$ MeV), as also shown by the distribution in the bottom left panel of Fig.~\ref{fg1}. For this reason, we adopt here the peak flux $P_{64}$ and fluence $F$ computed in the wide 10--1000 keV energy range as provided in the spectral catalogue of \fe\ bursts rather than the typically adopted 50--300 keV peak flux (e.g. from the \ba\ archive) which would sample only a portion of the full spectral curvature. 

The distributions of the peak flux, fluence, peak energy and duration are shown in Fig.~\ref{fg1} with black symbols. Error bars are computed by resampling each measurement ($P$, $F$, \epo\ and $T_{90}$) within its error with a normal distribution. For each bin, the vertical error bars represent the standard deviation of the bin heights of the resampled distributions.

\subsection{Rest-frame constraints: \textit{Swift} SBAT4 sample}

\begin{figure*}[htbp]
   \centering
    \vskip -3.4 truecm
    \hskip -0.8 truecm
\includegraphics[trim=1.3cm 0.3cm 0.55cm 0.2cm, clip,scale=0.48]{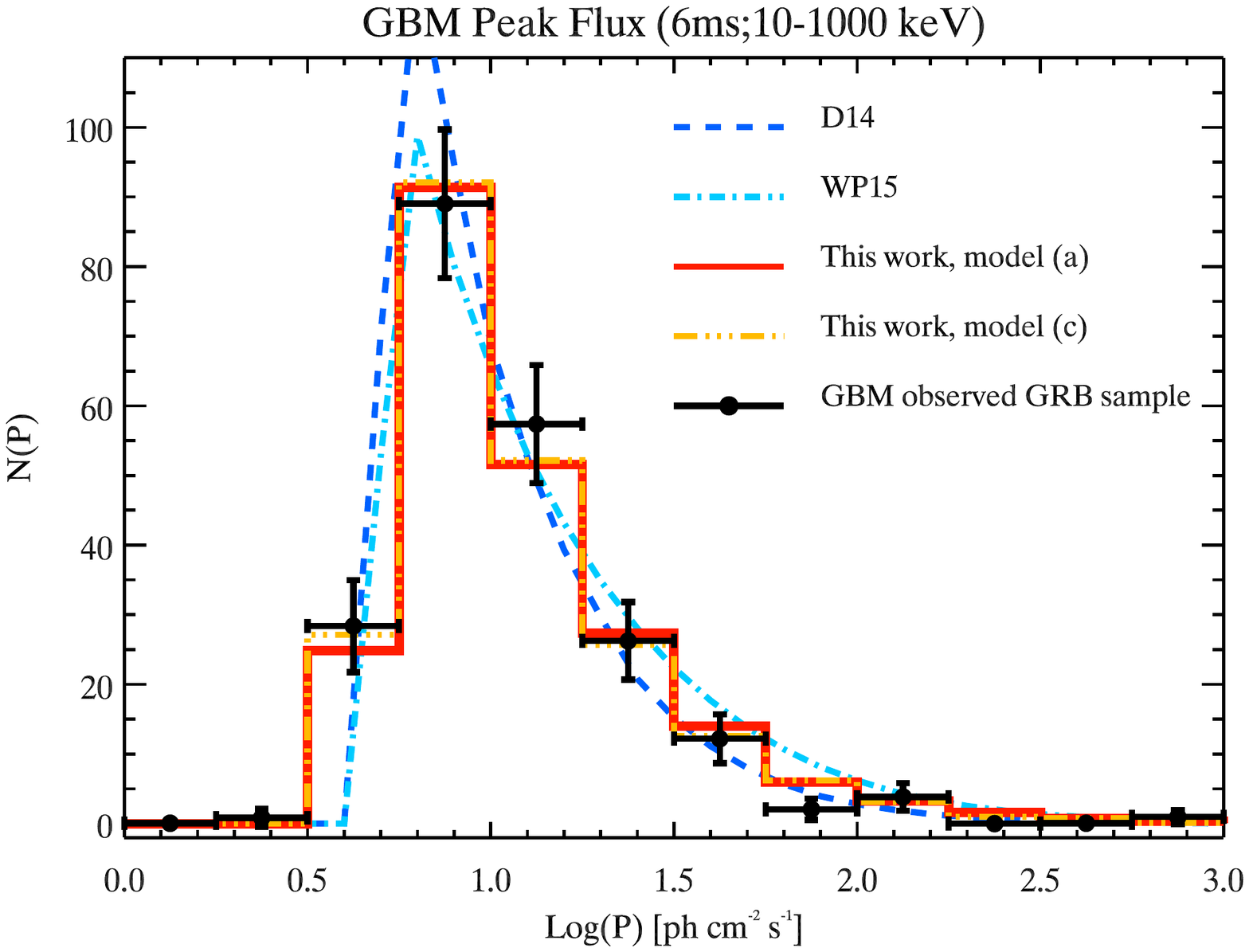} 
\includegraphics[trim=1.3cm 0.3cm 0.55cm 0.2cm, clip,scale=0.48]{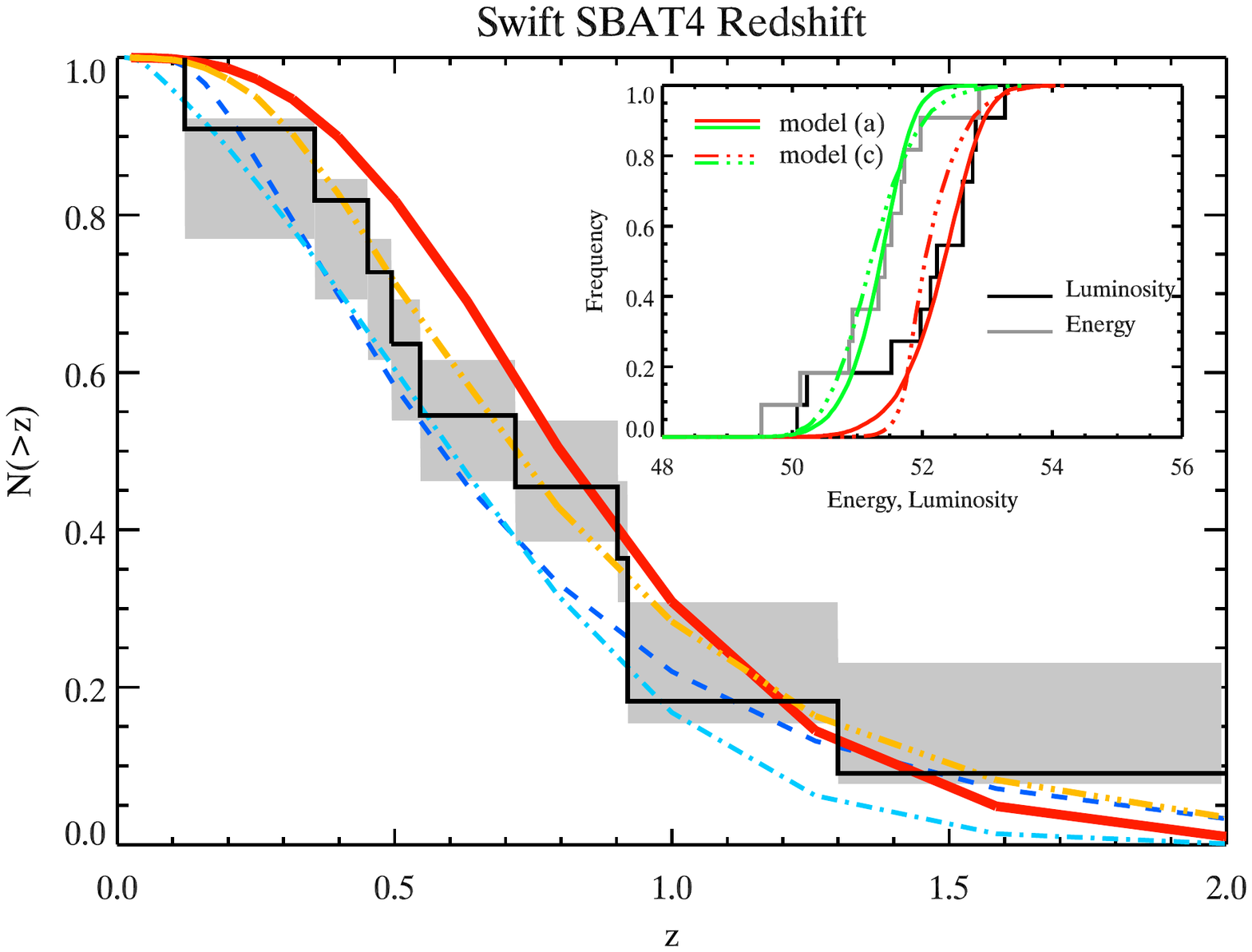} 
\vskip -6 truecm
\hskip -0.8 truecm
\includegraphics[trim=1.3cm 0.3cm 0.55cm 0.2cm, clip,scale=0.32]{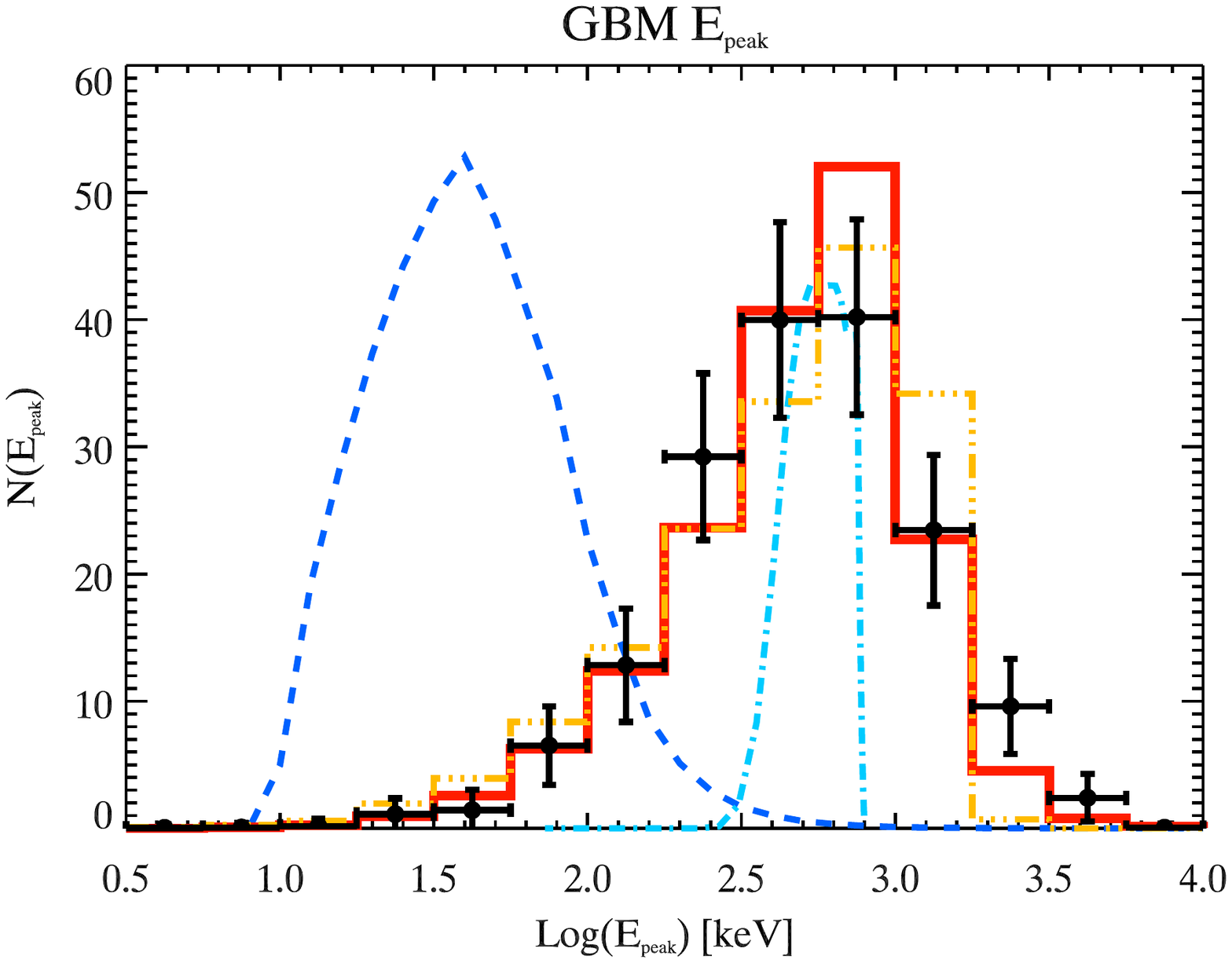} 
\includegraphics[trim=1.3cm 0.3cm 0.55cm 0.2cm, clip,scale=0.32]{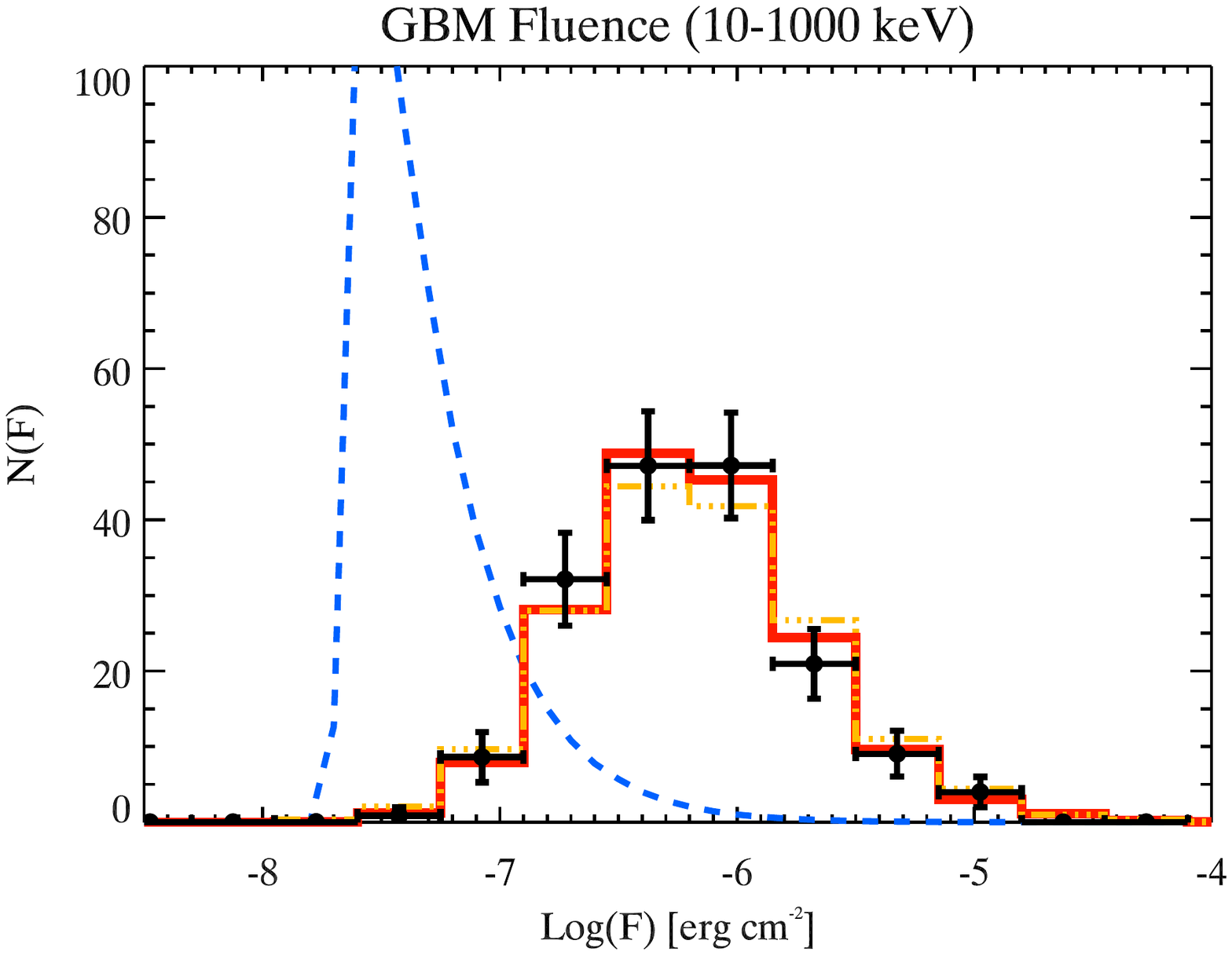} 
\includegraphics[trim=1.3cm 0.3cm 0.55cm 0.2cm, clip,scale=0.32]{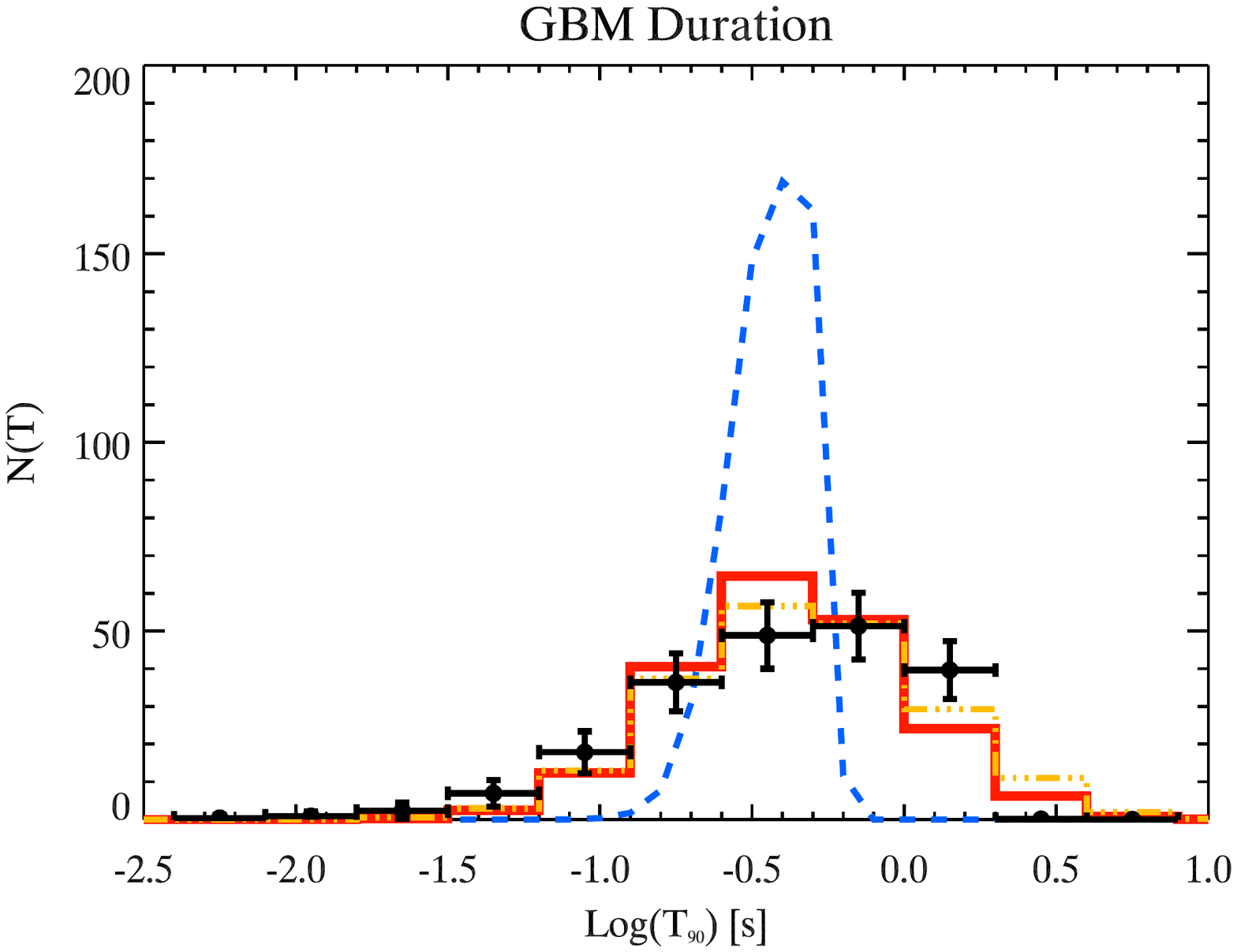} 
 \vskip -2.5 truecm
\caption{Black dots show the distributions obtained from our \textit{Fermi}/GBM and \sw\ SBAT4 samples (\S\ref{sec:sample_selection}). Horizontal error bars are the bin widths, while vertical error bars are 1 sigma errors on the bin heights accounting for experimental errors on single measurements. The results of our Monte Carlo population synthesis code are shown by solid red lines (assuming \yone\ and \ama\ correlations to hold in the population of SGRBs) and by triple dot--dashed orange lines (assuming no correlation). Predictions based on the models of D14 and WP15 are shown by dashed blue and dot--dashed cyan lines respectively (the latter only in the first three panels - see text). These are obtained by the analytical methods of \S\ref{sec:analytical_methods}. {\it Top left panel:} distribution of the peak flux $P$ of the \fe/GBM sample. {\it Top right panel:} normalised cumulative redshift distribution of the SBAT4 sample. The gray shaded area represents the range span by the distribution if the remaining bursts with unknown $z$ are assigned the largest or the lowest $z$ of the sample. The inset shows the cumulative distributions of the isotropic luminosity \liso\ (solid black line) and energy \eiso\ (grey solid line) of the same sample. {\it Bottom panels:} from left to right, distributions of peak energy \epo, fluence and duration of SGRBs of our \fe/GBM sample.
\label{fg1} }

\end{figure*}

For the redshift distribution and the rest frame properties of SGRBs (constraints 2, 6 and 7) we consider the sample published in D14. It consists of bursts detected by \textit{Swift}, selected with criteria similar to those adopted for the long GRBs in \cite{2012ApJ...749...68S},  with a peak flux (integrated in the 15--150 keV energy range and computed on a 64 ms timescale) $P_{64}\ge 3.5$ photons cm$^{-2}$ s$^{-1}$. This corresponds to a flux which is approximately 4 times larger than the \sw--BAT minimum detectable flux on this timescale. We call this sample SBAT4 (Short BAT 4) hereafter. The redshift distribution of the SBAT4 sample is shown in the top right panel of Fig.~\ref{fg1} (solid black line). Within the SBAT4 sample we consider the 11 GRBs with known $z$ and determined \liso\ and \eiso\ (the distributions of these quantities are shown in the inset of Fig.\ref{fg1}, top--right panel, with black and gray lines respectively). The gray shaded region is span by the distribution when the five SGRBs in the sample with unknown $z$ are all assigned the minimum or the maximum redshift of the sample. 


\section{The \lf\ and $\Psi(z)$ of SGRBs}\label{sec:LF_and_Psi}

Given the incompleteness of the available SGRB samples, particularly with measured $z$, no direct method \cite[alike for the population of long GRBs - see e.g.][]{2016A&A...587A..40P} can be applied to derive the shape of the SGRB luminosity function \lf\ and redshift distribution $\Psi(z)$ from the observations. The typical approach in this case consists in assuming some simple analytical shape for both functions, with free parameters to be determined by comparison of model predictions with observations. 

For the luminosity function, a power law 
\begin{equation}
\phi(L)\propto L^{-\alpha} 
\end{equation}
or a broken power law
\begin{equation}
\phi(L)\propto\left\lbrace\begin{array}{cc}
               \left({L}/{L_{\rm b}}\right)^{-\alpha_1} & L<L_{\rm b}\\
               \left({L}/{L_{\rm b}}\right)^{-\alpha_2} & L\geq L_{\rm b}
              \end{array}\right.
\end{equation}
normalised to its integral is usually assumed.

If SGRBs are produced by the merger of compact objects, their redshift distribution should follow a retarded star formation:
\begin{equation}
\Psi(z) = \int_{z}^{\infty} \psi(z')P[t(z)-t(z')]\frac{dt}{dz'}dz'
\label{eq:retarded}
\end{equation}
where \sf\ represents the formation rate of SGRB progenitors in Gpc$^{-3}$ yr$^{-1}$, and $P(\tau)$ is the delay time distribution,  i.e.\ the probability density function of the delay $\tau$ between the formation of the progenitors and their merger (which produces the SGRB). Adopting the point of view that SGRBs are produced by the coalescence of a neutron star binary (or a black hole -- neutron star binary), one can assume a delay time distribution and convolve it with a \sf\ of choice to obtain the corresponding SGRB formation rate $\Psi(z)$. Theoretical considerations and population synthesis \citep{1998A&A...332..173P,2001MNRAS.324..797S,2006ApJ...648.1110B,2008ApJ...675..566O,2013ApJ...779...72D} suggest that compact binary coalescences should typically follow a delay time distribution $P(\tau) \propto \tau^{-1}$ with $\tau\gtrsim 10$ Myr. 
Eq.~\ref{eq:retarded} is actually a simplification, in that it implicitly assumes that the fraction of newly formed stars that will end up as members of a NS--NS binary is fixed. The actual fraction very likely depends on metallicity and on the initial mass function, and thus on redshift in a statistical sense. 



Among the most recent studies of the \lf\ and \psiz\ of SGRBs we consider the work of D14 and WP15 in the following for comparison in more detail.
D14 assume a power law shape for both the \lf\ and the delay time distribution $P(\tau)$, and they adopt the parametric function of \cite{2001MNRAS.326..255C} for the cosmic star formation history, with parameter values from \cite{2006ApJ...651..142H}. They assume that SGRBs follow the \yone\ correlation $E_{\rm peak}=337 {\rm keV} (L_{\rm iso}/2\times 10^{52} {\rm erg s^{-1}})^{0.49}$ and that their spectrum is a Band function \citep{1993ApJ...413..281B} with low and high energy photon spectral indices -0.6 and -2.3, respectively. They constrain the free parameters by fitting the \ba\ peak flux distribution and the redshift distribution of bright \sw\ short bursts with measured $z$. They find $\phi(L)\propto L^{-2.17}$ between $10^{49}$ erg s$^{-1}$ and $10^{55}$ erg s$^{-1}$, and $P(\tau)\propto \tau^{-1.5}$ with a minimum delay of $20$ Myr. The dashed blue lines in Fig.~\ref{fg1} are obtained through Eq.~4 and Eq.~5 using the same parameters as D14: their model (limited to $P_{\rm lim}\ge 5$ \ph\ in order to be compared with the sample selected in this work) reproduces correctly the peak flux distribution (top left panel of Fig.~\ref{fg1}) of \fe\ SGRBs and the redshift distribution of the bright SGRBs detected by \sw\ (top right panel in Fig.~\ref{fg1}). 

The preferred model for $\phi(L)$ in WP15 is a broken power law, with a break at $2\times 10^{52}$ erg s$^{-1}$ and pre- and post break slopes of $-1.9$ and $-3.0$ respectively. Their preferred models are either a power law delay time distribution $P(\tau)\propto \tau^{-0.81}$ with a minimum delay of $20$ Myr or a lognormal delay time distribution with central value $2.9$ Gyr and sigma $\leq 0.2$. Differently from D14, rather than assuming the \yone\ correlation they assign to all SGRBs a fixed rest frame $E_{\rm p,rest}=800$ keV. The dot--dashed cyan lines in Fig.~\ref{fg1} are the model of WP15 (we show the lognormal $P(\tau)$ case).  

In the following we show how the results of WP15 and D14, both representative of a relatively steep luminosity function, compare with the other additional constraints (bottom panels of Fig.~\ref{fg1}) that we consider in this work. 


\subsection{From population properties to observables}\label{sec:analytical_methods}

Given the two functions \lf\ and $\Psi(z)$, the peak flux distribution can be derived as follows:
\begin{equation}
N(P_{1}<P<P_{2})=\frac{\Delta\Omega}{4 \pi}\int_{0}^{\infty}dz \frac{dV(z)}{dz}\frac{\Psi(z)}{1+z}\int_{L(P_{1},z)}^{L(P_{2},z)}\phi(L)dL
\label{eq:pfluxdistribution}
\end{equation}
where $\Delta\Omega/4\pi$ is the fraction of sky covered by the instrument/detector (which provides the real GRB population with which the model is to be compared) and $dV(z)/dz$ is the differential comoving volume.  The flux $P$ corresponding to the luminosity $L$ at redshift $z$ is\footnote{The assumption of a spectrum is required to convert the bolometric flux into a characteristic energy range for comparison with real bursts.}: 
\begin{equation}
P(L,z,E_{\rm peak},\alpha)=\frac{L}{4\pi d_{L}(z)^2}\, \frac{ \int_{\epsilon_{1}(1+z)}^{\epsilon_{2}(1+z)} N(E|E_{\rm peak},\alpha)dE}{\int_{0}^{\infty}EN(E|E_{\rm peak},\alpha)dE}
\end{equation}
where $d_{L}(z)$ is the luminosity distance at redshift $z$ and $N(E|E_{\rm peak},\alpha)$ is the rest frame photon spectrum of the GRB. The photon flux $P$ is computed in the rest frame energy range $[(1+z)\epsilon_{1},(1+z)\epsilon_{2}]$ which corresponds to the observer frame $[\epsilon_{1},\epsilon_{2}]$ band.

The SGRB spectrum is often assumed to be a cut-off power law, i.e. $N(E|E_{\rm peak},\alpha)\propto E^{-\alpha}\exp(-E(2-\alpha)/E_{\rm peak})$, or a Band function \citep{1993ApJ...413..281B}. Typical parameter values are  $\alpha\sim0.6$ \citep[i.e.~the central value of the population of SGRBs detected by \ba\ and \fe\ -][]{2009A&A...496..585G,2011A&A...530A..21N,2010int..workE..94G,2014ApJS..211...12G} and, for the Band function, $\beta\sim2.3-2.5$. The peak energy is either assumed fixed (e.g. $800$ keV in WP15) or derived assuming that SGRBs follow an \yone\ correlation in analogy to long bursts \citep[e.g.~D14;][]{2011ApJ...727..109V}. Recent evidence supports the existence of such a correlation among SGRBs \citep[see e.g.~D14;][]{2015MNRAS.448..403C,Tsutsui:2013lr,2009A&A...496..585G}, with similar parameters as that present in the population of long GRBs \citep{2004ApJ...609..935Y}. 

In order to compare the model peak flux distribution obtained from Eq.~\ref{eq:pfluxdistribution}  with the real population of GRBs, only events with peak flux above a certain threshold $P_{\rm lim}$ are considered. The integral in Eq.~\ref{eq:pfluxdistribution} is thus performed over the $(L,z)$ range where the corresponding flux is larger than $P_{\rm lim}$.

In D14 the assumption of the  correlation (\yone) between the isotropic luminosity \liso\ and the rest frame peak energy \ep\  allows us to derive, from Eq.~\ref{eq:pfluxdistribution}, also the expected distribution of the observer frame peak energy \epo: 
\begin{equation}
N(E_{1,{\rm p,o}}<E<E_{2, {\rm p,o}})=\int_{0}^{\infty}dz\,C(z)\int_{L(E_{1{\rm p,o}},z)}^{L(E_{2,{\rm p,o}},z)}\phi(L)dL
\label{eq:epdistribution}
\end{equation}
where $E_{\rm p,o}$ is the peak energy of the observed $\nu\,F(\nu)$ spectrum, and we let $C(z) = [\Delta\Omega/4\pi][\Psi(z)/(1+z)][dV(z)/dz]$. The limits of the luminosity integral are computed by using the rest frame correlation $E_{\rm p}=Y\,L^{m_y}$, namely
\begin{equation}
 L(E_{{\rm p,o}},z) = \left(\frac{E_{\rm p}}{Y}\right)^{1/m_y} = \left(\frac{(1+z)E_{\rm p,o}}{Y}\right)^{1/m_y}
\end{equation} 
In order to compare the distribution of \epo\ with real data, the integral in Eq.~\ref{eq:epdistribution}, similarly to Eq.~\ref{eq:pfluxdistribution}, is performed over values of $L(E_{\rm p, o},z)$ corresponding to fluxes above the limiting flux adopted to extract the real GRB sample (e.g. 5 \ph\ for SGRBs selected from the \fe\ sample). 

Similarly, by assuming an \ama\ correlation to hold in SGRBs \citep[see D14;][]{Tsutsui:2013lr,2006MNRAS.372..233A,2015MNRAS.448..403C}, i.e.\ $E_{\rm p}=A\,E^{m_a}$, we can derive a relation between luminosity and energy (\liso--\eiso), which reads
\begin{equation}
 L(E) = \left(\frac{A}{Y}\right)^{1/m_y} E^{m_a/m_y}
\end{equation} 
This is then used to compute the fluence distribution, where the fluence is related to the isotropic energy as $F=E(1+z)/4\pi\,d_{L}(z)^{2}$:
\begin{equation}
N(F_{1}<F<F_{2})=\int_{0}^{\infty}dz\,C(z)\int_{L(E_{1})}^{L(E_{2})}\phi(L)dL
\end{equation}
again by limiting the integral to luminosities which correspond to fluxes above the given limiting flux.

Finally, considering the spiky light curves of SGRBs, we can assume a triangular shape and thus let $2E/L\sim T$ in the rest frame of the source. Therefore, it is possible to combine the \ama\ and \yone\ correlations to derive the model predictions for the distribution of the duration to be compared with the observed one:
\begin{equation}
N(T_{1,\rm o}<T<T_{2,\rm o})=\int_{0}^{\infty}dz\,C(z)\int_{L(T_{1,{\rm o}},z)}^{L(T_{2,{\rm o}},z)}\phi(L)dL
\end{equation}
where 
\begin{equation}
 L(T_{\rm o},z) = \left[\left( \frac{Y}{A} \right)^{1/m_a}\frac{2(1+z)}{T_o}\right]^{1/(1-m_y/m_a)}
\end{equation} 

\subsection{Excluding a steep luminosity function}

The bottom panels of Fig.~\ref{fg1} show the distributions of peak energy \epo\ (left), fluence $F$ (middle) and duration $T_{90}$ (right) of the sample of short \fe\ GRBs described in \S2 (black symbols). Predictions using the same parameters as in D14 are shown by dashed blue lines in Fig.~\ref{fg1}: while the $P$ and $z$ distributions are correctly reproduced (top panels of Fig.~\ref{fg1}), the model is inconsistent with the distributions of peak energy \epo, fluence $F$  and  duration (bottom panels of Fig.~\ref{fg1}). For the D14 model we have assumed the \ama\ correlation reported in that paper to derive the fluence and (in combination with the \yone\ correlation) the duration distribution. Since WP15 assume a unique value of the peak energy \epo\ it is not possible to derive the fluence and duration of their model, unless independent functions for these parameters are assumed. Therefore, the model of WP15 (dot--dashed cyan line in Fig.~\ref{fg1}) is compared only in the peak flux, redshift (top panels) and observed peak energy (bottom left panel of Fig.~\ref{fg1}).

In conclusion, a steep \lf\ with either a power law distribution of delay times favoring short delays (as in D14) or a nearly unique long delay time (as in the log--normal model of WP15) correctly reproduce the observer frame peak flux distribution of \fe\ GRBs\footnote{Here we consider as constrain the population of \fe/GBM GRBs. \cite{2011MNRAS.415.3153N} showed that that the \ba\ SGRB population has similar prompt emission  (peak flux, fluence and duration distribution) properties of \fe\ SGRBs.} and the redshift distribution of \sw\ bright short bursts. However, they do not reproduce the peak energy, fluence and duration distributions of the same population of \fe\ SGRBs.


Motivated by these results, we implemented a Monte Carlo (MC) code aimed at deriving the \lf\ and \psiz\ of SGRBs which satisfy all the constraints (1--7) described above. The reason to choose a MC method is that it allows to easily implement the dispersion of the correlations (e.g. \yone\ and \ama) and of any distribution assumed (which are less trivial to account for in an analytic approach as that shown above).

\section{Monte Carlo simulation of the SGRB population}

\begin{figure}
\centering
\vskip -1.8 truecm 
 \includegraphics[width=0.9\columnwidth]{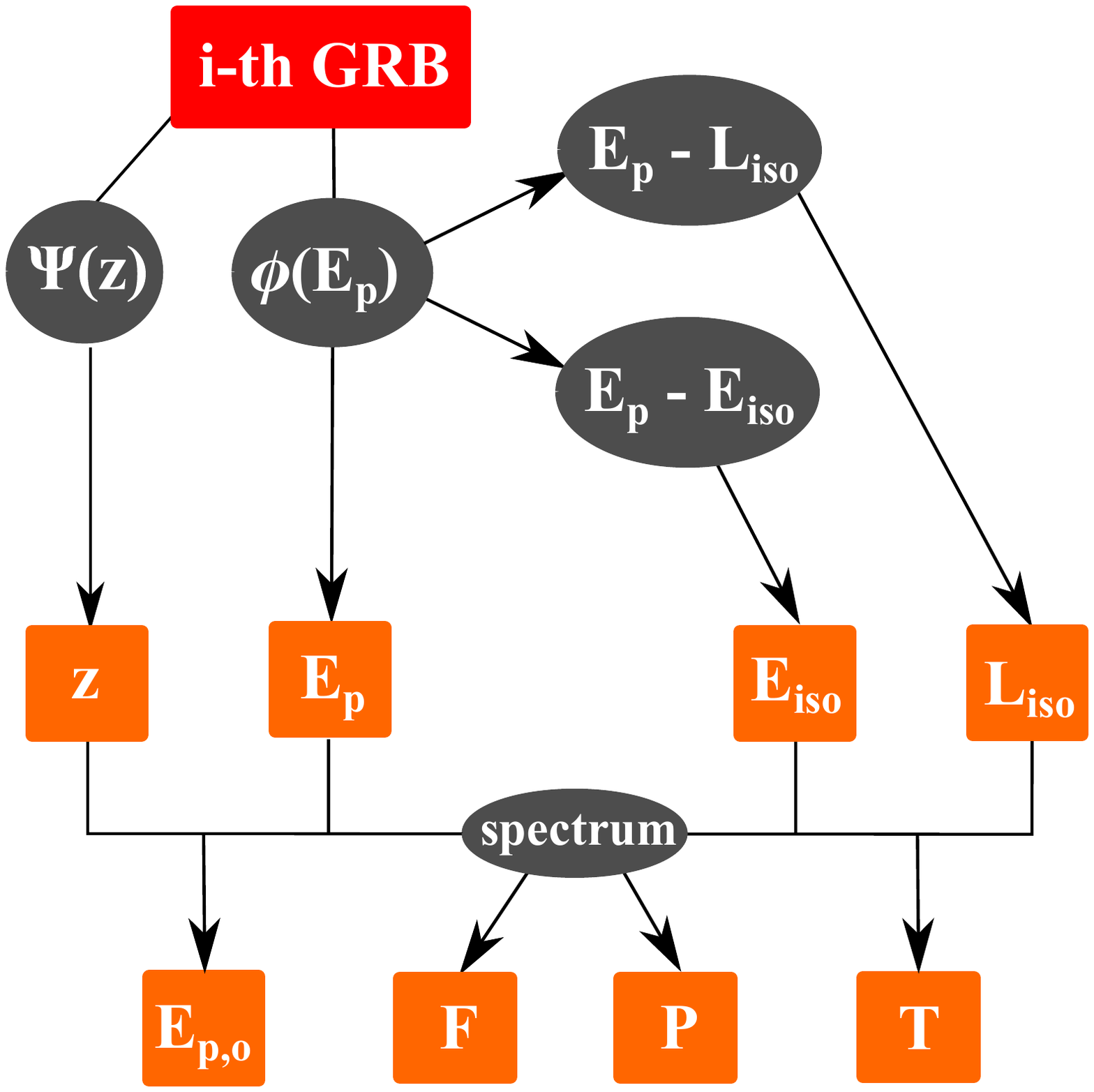}
 \vskip -2.5 truecm
\caption{\label{fig:MCscheme}Scheme of the procedure followed in the MC to generate the observables of each synthetic GRB.}
\end{figure}

In this section we describe the Monte Carlo (MC) code adopted to generate the model population. Such population is then compared with the real SGRB samples described above in order to constrain the model parameters (\S4). Our approach is based on the following choices:

\begin{enumerate}
\item

Customarily, Eq.~\ref{eq:retarded} has been used to compute the redshift distribution $\Psi(z)$ of SGRBs from an assumed star formation history $\psi(z)$ and a delay time distribution $P(\tau)$. As stated in \S\ref{sec:LF_and_Psi}, this approach implies simplifications that we would like to avoid. To make our analysis as general as possible, we here adopt a generic parametric form for the redshift distribution $\Psi(z)$ of SGRBs. \textit{A posteriori}, if one believes the progenitors to be compact binaries, the delay time distribution can be recovered by direct comparison of our result with the star formation history of choice. We parametrise the \psiz\ following \cite{2001MNRAS.326..255C}, namely:
\begin{equation}
\Psi(z) = \frac{1+p_{1}z}{1+\left(z/z_{\rm p}\right)^{p_{2}}}
\label{eq:psi_cole}
\end{equation}
which has a rising and decaying part (for $p_{1}>0$, $p_{2}>1$) and a characteristic peak roughly\footnote{The exact peak is not analytical, but a good approximation is $z_{\rm{peak}} \approx z_{\rm{p}}\left\lbrace p_2\left[1 + 1/\left(p_1\,z_{\rm{p}}\right)\right]-1\right\rbrace^{-1/p_2}$.} corresponding to $z_{p}$;

\item In order to have a proper set of simulated GRB parameters, it is convenient to extract $E_{\rm p}$ from an assumed probability distribution. We consider a broken power law shape for the \ep\ distribution: 
\begin{equation}
\phi(E_{\rm p}) \propto
\begin{cases}
\left({E_{p}}/{E_{\rm p,b}}\right)^{-a_1}  & \text{ $E_{p} \leq E_{\rm p,b}$} \\
\left({E_{p}}/{E_{\rm p,b}}\right)^{-a_2}   & \text{ $E_{p}> E_{\rm p,b}$}
\end{cases}
\label{lf}
\end{equation}
Through the \yone\ and \ama\ correlations, accounting also for their scatter, we can then associate to $E_{\rm p}$ a luminosity \liso\ and an energy \eiso. The luminosity function of the population is then constructed as a result of this procedure;

\item We assume the \yone\ and \ama\ correlations to exist and we write them respectively as 
\begin{equation}
 \log_{10}(E_{\rm p}/670\,{\rm keV}) = q_{\rm{Y}} + m_{\rm{Y}}\log_{10}(L/10^{52}\rm{erg\,s^{-1}})
 \label{eq:yone}
\end{equation}  
and 
\begin{equation}
 \log_{10}(E_{\rm p}/670\,{\rm keV}) = q_{\rm{A}} + m_{\rm{A}}\log_{10}(E_{\rm iso}/10^{51}\rm{erg})
 \label{eq:ama}
\end{equation}
After sampling $E_{\rm p}$ from its probability distribution (Eq.~\ref{lf}), we associate to it a luminosity (resp. energy) sampled from a lognormal distribution whose central value is given by Eq.~14 (resp. 15) and $\sigma=0.2$. The SGRBs with measured redshift are still too few to measure the scatter of the corresponding correlations. We assume the same scatter as measured for the correlations holding for the population of long GRBs \citep{Nava:2012lr};

\item For each GRB, a typical Band function prompt emission spectrum is assumed, with low and high photon spectral index $-0.6$ and $-2.5$ respectively. We keep these two parameters fixed after checking that our results are unaffected by sampling them from distributions centred around these values\footnote{We also tested that our results are not sensitive to a slightly different choice of the spectral parameters, i.e. low and high energy spectral index $-1.0$ and $-3.0$ respectively.}. 
\end{enumerate}

For each synthetic GRB, the scheme in Fig.~\ref{fig:MCscheme} is followed: a redshift $z$ is sampled from $\Psi(z)$ and a rest frame peak energy \ep\ is sampled from $\phi(E_{\rm p})$; through the \yone\ (\ama) correlation a luminosity \liso\ (energy \eiso) is assigned, with lognormal scatter; using redshift and luminosity (energy), via the assumed spectral shape, the peak flux $P$ (fluence $F$) in the observer frame energy range 10--1000 keV is derived. The observer frame duration $T$ is obtained as $2(1+z)E/L$, i.e.\ the light curve is approximated with a triangle\footnote{This might seem a rough assumption, since SGRBs sometimes show multi peaked light curves. Statistical studies, however, show that the majority of SGRB lightcurves are composed of few peaks, with separation much smaller than the average duration (e.g. \cite{McBreen:2001fk}), which justifies the use of this assumption in a statistical sense.}. Let us refer to this scheme as ``case (a)''.

\label{sec:caseb}
The minimum and maximum values of $E_{\rm p}$ admitted are $E_{\rm p,min} = 0.1\,\rm{keV}$ and $E_{\rm p,max} = 10^5\,\rm{keV}$. These limiting values correspond to a minimum luminosity $L_{\rm min}$ and a maximum luminosity $L_{\rm max}$ which depend on the \yone\ correlation. While the maximum luminosity is inessential (in all our solutions the high luminosity slope $\alpha_2 \gtrsim 2$), the existence of a minimum luminosity might affect the observed distributions. We thus implemented an alternative scheme (``case (b)'') where the minimum luminosity $L_{\rm min}$ is a parameter, and values of $E_{\rm p}$ which correspond to smaller luminosities are rejected.

In order to investigate the dependence of our results on the assumption of the \yone\ and \ama\ correlations, we also implemented a third MC scheme (``case (c)'') where independent (from the peak energy and between themselves) probability distributions are assumed for the luminosity and duration. A broken power law
\begin{equation}
 P(L) \propto \left\lbrace\begin{array}{lr}
                              (L/L_{\rm b})^{-\alpha_1} & L\leq L_{\rm b}\\
                              (L/L_{\rm b})^{-\alpha_2} & L>L_{\rm b}\\
                             \end{array}\right.\label{eq:lf}
\end{equation} 
is assumed for the luminosity distribution, and a lognormal shape
\begin{equation}
 P(T_{\rm{r}}) \propto \exp\left[-\frac{1}{2}\left(\frac{(\log(T_{\rm{r}}) - \log(T_c)}{\sigma_{Tc}}\right)^2\right]
 \label{eq:tdist}
\end{equation}
is assumed for the rest frame duration $T_{\rm{r}} = T/(1+z)$ probability distribution. Again, the energy of each GRB is computed as $E = L T_{\rm{r}}/2$, i.e.\ the light curve is approximated with a triangle.

\section{Finding the best fit parameters}

In case (a) there are 10 free parameters: three $(p_{1},z_{\rm p},p_{2})$ define the redshift distribution (Eq.~\ref{eq:psi_cole}), three $(a_1,a_2,E_{\rm p,b})$ define the peak energy distribution (Eq.~\ref{lf}) and four $(q_Y,m_Y,q_A,m_A)$ define the \yone\ and \ama\ correlations (Eqs.~\ref{eq:yone}\,\&\ref{eq:ama}). Our constraints are the seven distributions shown in Fig.~\ref{fg1} (including the top right panel insets). 

\begin{figure*}
\centering
\includegraphics[width=\textwidth]{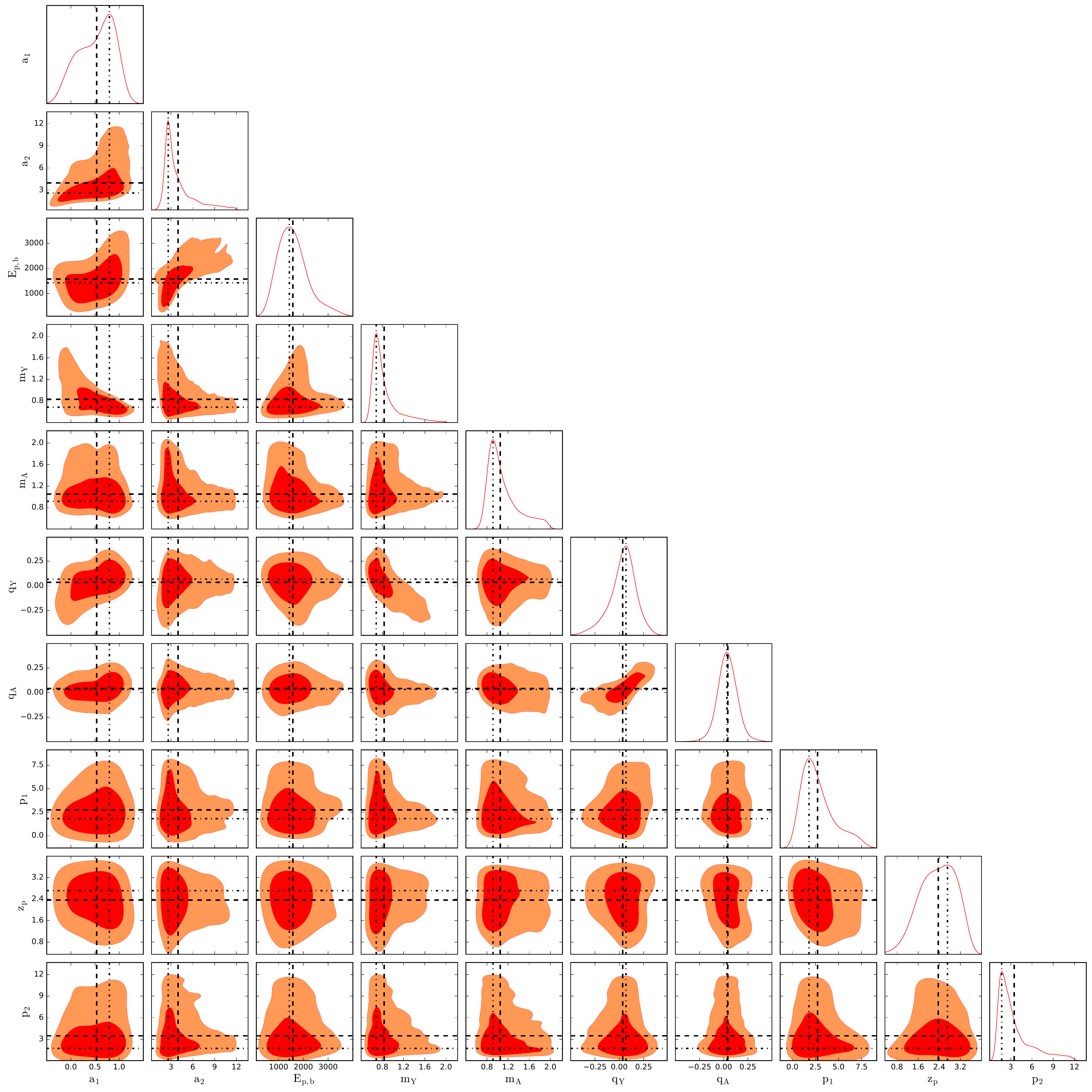}
\caption{Marginalized densities of our MCMC parameters in ``case (a)'' (i.e. with correlations and no minimum luminosity). Black dashed lines indicate the means and black dot-dashed lines indicate the modes of the distributions.\label{fig:triangolo}}
 \end{figure*}

In order to find the best fit values and confidence intervals of our parameters, we employed a Monte Carlo Markov Chain (MCMC) approach based on the Metropolis-Hastings algorithm \citep{hastings1970monte}. At each step of the MCMC: 
\begin{itemize}
 \item we displace each parameter\footnote{For parameters corresponding to slopes, like $m_{\rm{Y}}$ and $m_{\rm{A}}$, we actually displace the corresponding angle $\phi=\arctan(m)$, otherwise a uniform sampling of the displacement would introduce a bias towards high (i.e.~steep) slopes.}  $p_i$ from the last accepted value. The displacement is sampled from a uniform distribution whose maximum width is carefully tuned in order to avoid the random walk to remain stuck into local maxima;
 \item we compute the Kolmogorov-Smirnov (KS) probability $P_{\rm KS,j}$ of each observed distribution to be drawn from the corresponding model distribution;
 \item we define the goodness of fit $\mathcal{G}$ of the model as the sum of the logarithms of these KS probabilities\footnote{This is clearly only an approximate likelihood, since it implies an assumption of independence of each distribution from the others, but we tested that its maximisation gives consistent results.}, i.e.\ \mbox{$\mathcal{G} = \sum_{j=1}^{7} \log P_{\rm KS,j}$};
 \item we compare $g=\exp({\mathcal{G}})$ with a random number $r$ sampled from a uniform distribution within $0$ and $1$: if $g>r$ the set of parameters is ``accepted'', otherwise it is ``rejected''.
\end{itemize}

We performed tests of the MCMC with different initial parameters, to verify that a unique global maximum of $\mathcal{G}$ could be found. Once properly set up, 200,000 steps of the MCMC were run. After removing the initial burn in, the autocorrelation length of each parameter in the chain was computed, and the posterior density distribution of each parameter (and the joint distribution of each couple of parameters) was extracted with the \texttt{getDist} python package\footnote{\texttt{getDist} is a python package written by Antony Lewis of the University of Sussex. It is a set of tools to analyse MCMC chains and to extract posterior density distributions using Kernel Density Estimation (KDE) techniques. Details can be found at \url{http://cosmologist.info/notes/GetDist.pdf}.}. The resulting 1D and 2D marginalized distributions are shown in Fig.~\ref{fig:triangolo}, where black dashed (black dot-dashed) lines indicate the position of the mean (mode) of the marginalized density of each parameter. The filled contours represent the 68\% (darker red) and 95\% (lighter red) probability areas of the joint density distributions. The means, modes and $68\%$ probability intervals of the 1D marginalized distributions are summarised in Table~\ref{tab:mcmc_results}.a, where the corresponding luminosity function parameters are also reported. 

\begin{table}
\caption{Summary of Monte Carlo Markov Chain results. C.I. = confidence interval. $E_{\rm peak,b}$, $L_b$ and $T_c$ are in units of keV, $10^{52}$ erg s$^{-1}$ and s, respectively. \label{tab:mcmc_results}}

\textbf{(a) case with correlations and no minimum luminosity}
\begin{center}

\begin{tabular}{llll}
  Parameter & Mean & Mode & $68\%$ C.I. \\
  \hline\hline
$a_1$ & $0.53$ & $0.8$ & $(0.2,1)$ \\
$a_2$ & $4$ & $2.6$ & $(1.9,4.4)$ \\
$E_{\rm peak,b}$ & $1600$ & $1400$ & $(880,2000)$ \\
$m_Y$ & $0.84$ & $0.69$ & $(0.58,0.88)$ \\
$m_A$ & $1.1$ & $0.91$ & $(0.76,1.2)$ \\
$q_Y$ & $0.034$ & $0.068$ & $(-0.069,0.18)$ \\
$q_A$ & $0.042$ & $0.033$ & $(-0.061,0.13)$ \\
$p_1$ & $2.8$ & $1.8$ & $(0.59,3.7)$ \\
$z_p$ & $2.3$ & $2.7$ & $(1.7,3.2)$ \\
$p_2$ & $3.5$ & $1.7$ & $(0.94,4)$ \\
\hline
$\alpha_1$ & $0.53$ & $0.88$ & $(0.39,1.0)$ \\
$\alpha_2$ & $3.4$ & $2.2$ & $(1.7,3.7)$ \\
$L_b$ & $2.8$ & $2.1$ & $(0.91,3.4)$ \\
\hline
\end{tabular}
 
\end{center}

\vspace{10pt}
\textbf{(b) case with correlations and minimum luminosity}

\begin{center}
 \begin{tabular}{llll}
  Parameter & Mean & Mode & $68\%$ C.I. \\
  \hline\hline
$a_1$ & $0.39$ & $0.24$ & $(-0.15,0.8)$ \\
$a_2$ & $3.5$ & $2.5$ & $(1.9,3.7)$ \\
$E_{\rm peak,b}$ & $1400$ & $1100$ & $(730,1700)$ \\
$m_Y$ & $0.88$ & $0.76$ & $(0.61,0.97)$ \\
$m_A$ & $1.1$ & $0.95$ & $(0.77,1.2)$ \\
$q_Y$ & $0.045$ & $0.077$ & $(-0.039,0.17)$ \\
$q_A$ & $0.043$ & $0.053$ & $(-0.037,0.14)$ \\
$p_1$ & $3.1$ & $2.4$ & $(1,4.2)$ \\
$z_p$ & $2.5$ & $3$ & $(1.9,3.3)$ \\
$p_2$ & $3$ & $1.3$ & $(0.9,3.1)$ \\
\hline
$\alpha_1$ & $0.38$ & $0.47$ & $(0.034,0.98)$ \\
$\alpha_2$ & $3$ & $2.1$ & $(1.7,3.2)$ \\
$L_b$ & $2.3$ & $1.5$ & $(0.71,2.8)$ \\
\hline
 \end{tabular}
 \end{center}

\vspace{10pt}
\textbf{(c) case with no correlations}

\begin{center}
 \begin{tabular}{llll}
  Parameter & Mean & Mode & $68\%$ C.I. \\
  \hline\hline
$a_1$ & $-0.61$ & $-0.55$ & $(-0.73,-0.41)$ \\
$a_2$ & $2.8$ & $2.5$ & $(2.1,2.9)$ \\
$E_{\rm peak,b}$ & $2200$ & $2100$ & $(1900,2500)$ \\
$\alpha_1$ & $-0.15$ & $-0.32$ & $(-1.5,0.81)$ \\
$\alpha_2$ & $2.0$ & $1.8$ & $(1.2,2.8)$ \\
$L_b$ & $0.63$ & $0.79$ & $(0.32,1.6)$ \\
$T_c$ & $0.11$ & $0.11$ & $(0.084,0.13)$ \\
$\sigma_{\rm{Tc}}$ & $0.91$ & $0.90$ & $(0.79,1.0)$ \\
$p_1$ & $3.1$ & $2.0$ & $(0.51,4.1)$ \\
$z_p$ & $2.5$ & $2.8$ & $(2.0,3.3)$ \\
$p_2$ & $3.6$ & $2.0$ & $(1.1,3.7)$ \\
\hline
\end{tabular}
\end{center} 
\end{table}

For the solution represented by the mean values in Table~\ref{tab:mcmc_results}.a, the minimum luminosity is $L_{\rm min}\sim 10^{47}$ erg s$^{-1}$. For comparison, we tested case (b) fixing $L_{\rm min}=10^{50}$ erg s$^{-1}$. This is the highest minimum luminosity one might assume, since the lowest SGRB measured luminosity in the \sw\ sample considered is  $L=1.2\times 10^{50}$ erg s$^{-1}$ (D14).  Table~\ref{tab:mcmc_results}.b summarises the results of the analysis after 200,000 MCMC steps. The two cases are consistent within one sigma. The best fit luminosity function in case (b) is slightly shallower at low luminosities (i.e.\ there is a slight decrease in $\alpha_1$) than in case (a), and it remains much shallower than in D14 and WP15. 

Finally, we tested case (c) performing 200,000 MCMC steps. In this case, the free parameters are eleven: three $(p_{1},z_{\rm p},p_{2})$ for $\Psi(z)$ and three $(a_1,a_2,E_{\rm p,b})$ for $\phi(E_{\rm p})$ as before, plus three $(\alpha_1,\alpha_2,L_{\rm b})$ for the luminosity function (Eq.~\ref{eq:lf}) and two $(T_c,\sigma_{\rm Tc})$ for the intrinsic duration distribution (Eq.~\ref{eq:tdist}). Consistently with case (a) and case (b) we assumed two broken power laws for $\phi(E_{\rm p})$ and \lf. Results are listed in Table~\ref{tab:mcmc_results}.c. We find that if no correlations are present between the peak energy and the luminosity (energy), the luminosity function and the peak energy distributions become peaked around characteristic values. This result is reminiscent of the findings of \cite{2015MNRAS.451..126S} who assumed lognormal distributions for these quantities.

\section{Discussion of the results}

\subsection{Luminosity function}

In case (a) we find that the luminosity function is shallow ($\alpha_1 = 0.53^{+0.47}_{-0.14}$ - and flatter than 1.0 within the 68\% confidence interval) below a break luminosity $\sim 3 \times 10^{52}$ erg s$^{-1}$ and steeper ($\alpha_2=3.4^{+0.3}_{-1.7}$) above this characteristic luminosity. The minimum luminosity $\sim 5\times 10^{47}$ erg s$^{-1}$ is set by the minimum $E_{\rm p}$ coupled with the \yone\ correlation parameters (see \S\ref{sec:caseb}). Similar parameters for the \lf\ are obtained in case (b), where a minimum luminosity was introduced, thus showing that this result is not strongly dependent on the choice of the minimum luminosity of the \lf.

If we leave out the correlations (case (c)), we find that the distributions of the peak energy and luminosity are peaked. However, the 68\% confidence intervals of some parameters, common to case (a) and (b), are larger in case (c). In particular, the slope $\alpha_1$  of the luminosity function below the break is poorly constrained, despite this cannot be steeper than 0.81 (at the 68\% confidence level). We believe that the larger uncertainty on the best fit parameters in case (c) is due to the higher freedom allowed by the uncorrelated luminosity function, peak energy distribution and duration distribution.

\subsection{Redshift distribution}

\begin{figure}
\begin{center}
 \includegraphics[width=0.9\columnwidth]{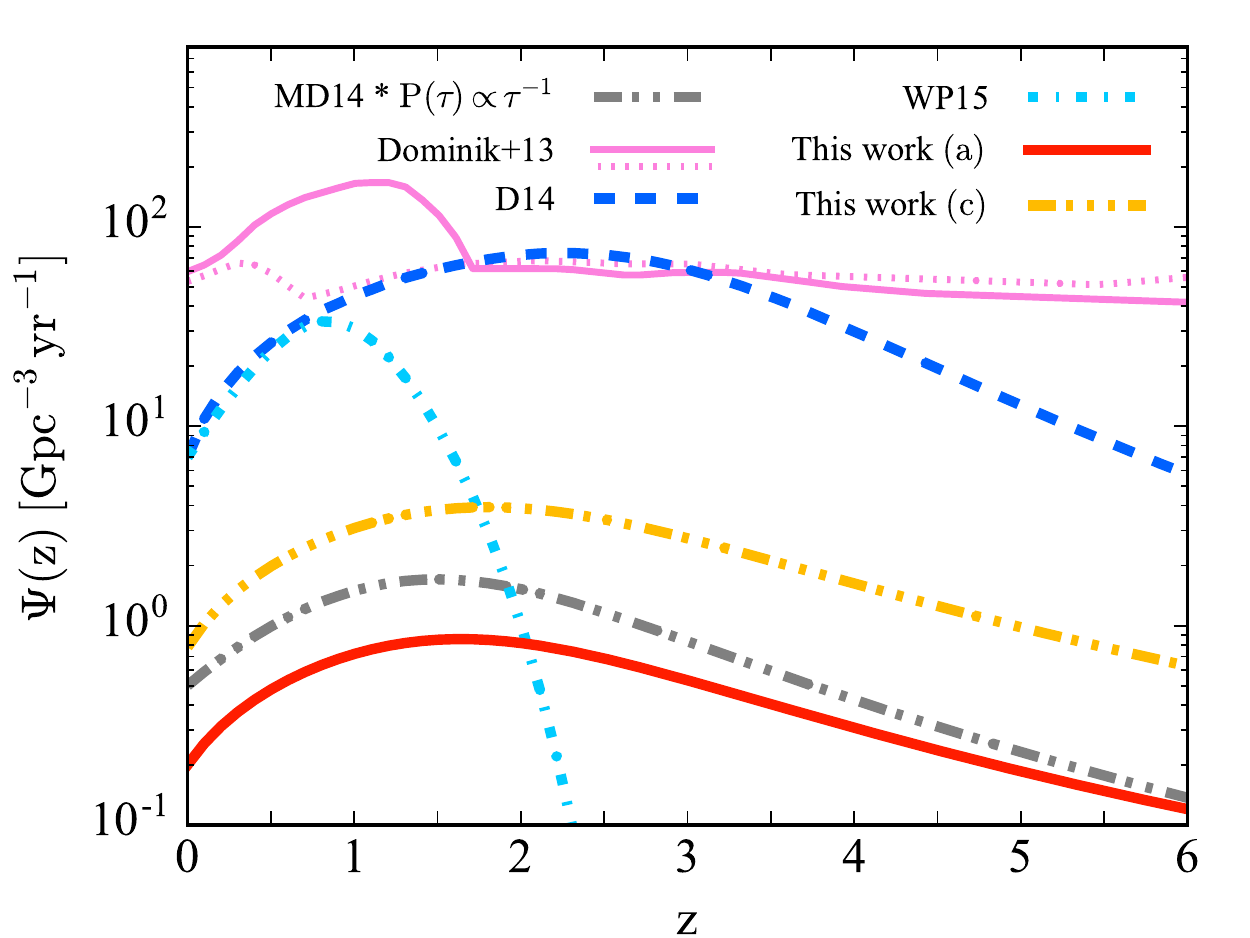}
\end{center}
\caption{Comparison between various predicted SGRB redshift distributions. The grey dashed line represents the convolution of the MD14 cosmic SFH with a delay time distribution $P(\tau)\propto \tau^{-1}$ with $\tau>20 \rm{Myr}$ (the normalization is arbitrary). The pink solid line (pink dotted line) represents the redshift distribution of NS--NS binary mergers predicted by \cite{2013ApJ...779...72D} in their \textit{high end} (\textit{low end}) metallicity evolution scenario (standard binary evolution model). The blue dashed line and cyan dot--dashed line are the SGRB redshift distributions according to D14 and to WP15 respectively. The red solid line is our result in case (a), while the orange triple dot dashed line is our result in case (c). In both cases we used the mean parameter values as listed in Table~\ref{tab:mcmc_results}.\label{fig:sfh_comparison}}
\end{figure}

Figure~\ref{fig:sfh_comparison} shows a comparison of our predicted redshift distributions (case (a): red solid line; case (c): orange triple dot-dashed line; mean values adopted) with the following other redshift distributions:
\begin{itemize}
 \item the convolution of the \cite[][MD14 hereafter]{2014ARA&A..52..415M} star formation history (SFH) with the delay time distribution $P(\tau)\propto \tau^{-1}$ with $\tau> 20\rm{Myr}$, grey dashed line (the normalisation is arbitrary);
 \item the redshift distribution of NS--NS mergers as predicted by \cite{2013ApJ...779...72D} (we refer to the standard binary evolution case in the paper) based on sophisticated binary population synthesis, assuming two different metallicity evolution scenarios: \textit{high-end} (pink solid line) and \textit{low-end} (pink dotted line);
 \item the SGRB redshift distribution found by D14, which is obtained convolving the SFH by \cite{2006ApJ...651..142H} with a delay time distribution $P(\tau)\propto \tau^{-1.5}$ with $\tau> 20\rm{Myr}$, blue dashed line;
 \item the SGRB redshift distribution found by WP15, which is obtained convolving an SFH based on Planck results \citep[``extended halo model'' in][]{2014A&A...571A..30P} with a lognormal delay time distribution $P(\tau)\propto \exp\left[-\left(\ln \tau - \ln \tau_0\right)^2/\left(2\sigma^2\right) \right]$ with $\tau_0 = 2.9\rm{Gyr}$ and $\sigma < 0.2$ (we used $\sigma = 0.1$), cyan dot--dashed line.
\end{itemize}

The redshift distribution by D14 peaks between $z\sim 2$ and $z\sim 2.5$, i.e.~at a higher redshift than the MD14 SFH (which peaks at $z\sim 1.9$). This is due to the short delay implied by the delay time distribution assumed in D14, together with the fact that the \cite{2006ApJ...651..142H} SFH peaks at higher redshift than the MD14 SFH. On the other hand, the redshift distribution by WP15 peaks at very low redshift ($\sim 0.8$) and predicts essentially no SGRBs with redshift $z\gtrsim 2$, because of the extremely large delay implied by their delay time distribution. 

Assuming the MD14 SFH (which is the most up-to-date SFH available) to be representative, our result in case (a) seems to be compatible with the $P(\tau)\propto \tau^{-1}$ delay time distribution (grey dashed line), theoretically favoured for compact binary mergers. In case (c), on the other hand, the redshift distribution we find seems to be indicative of a slightly smaller average delay with respect to case (a). Since the cosmic SFH is still subject to some uncertainty, and since the errors on our parameters $(p_1,z_p,p_2)$ are rather large, though, no strong conclusion about the details of the delay time distribution can be drawn.

\subsection{\yone\ and \ama\ correlations}

Our approach allowed us, in cases (a) and (b), to derive the slope and normalization of the intrinsic \yone\ and \ama\ correlations of SGRBs. 
\cite{Tsutsui:2013lr} finds, for the \ama\ and \yone\ correlations of SGRBs, slope values $0.63\pm0.05$ and $0.63\pm0.12$, respectively. Despite our mean values for $m_{Y}$ and $m_{A}$ (Tab.~1) are slightly steeper, the 68\% confidence intervals reported in Tab.~1 are consistent with those reported by  \cite{Tsutsui:2013lr}. In order to limit the free parameter space we assumed a fixed scatter for the correlations and a fixed normalisation center for both (see Eq.~14 and Eq.~15). This latter choice, for instance, introduces the small residual correlation between the slope and normalisation of the \yone\ parameters (as shown in Fig.~\ref{fig:triangolo}). 

Inspection of Fig.~\ref{fig:triangolo} reveals another correlation in the MCMC chain between the normalizations $q_{\rm{Y}}$ and $q_{\rm{A}}$ of the \yone\ and \ama\ correlations: this is expected, because the ratio of the two normalizations is linked to the duration of the burst. Indeed, from Eqs.~\ref{eq:ama} \& \ref{eq:yone} one has
\begin{equation}
 q_{\rm{Y}} - q_{\rm{A}} = \log\left(\frac{E^{m_{\rm{A}}}}{L^{m_{\rm{Y}}}}\right) + 52 m_{\rm{Y}} - 51 m_{\rm{A}}
\end{equation}
Since $m_{\rm{A}}$ and $m_{\rm{Y}}$ are close, the argument of the logarithm is $\sim E/L \propto T$, and since there is a typical duration, this induces an approximately linear correlation between $q_{\rm{A}}$ and $q_{\rm{Y}}$, which is what we find.

\begin{figure*}[htbp]
\centering
\vskip -6.8 truecm
\hskip -1.4truecm
\centering
\includegraphics[width=\textwidth]{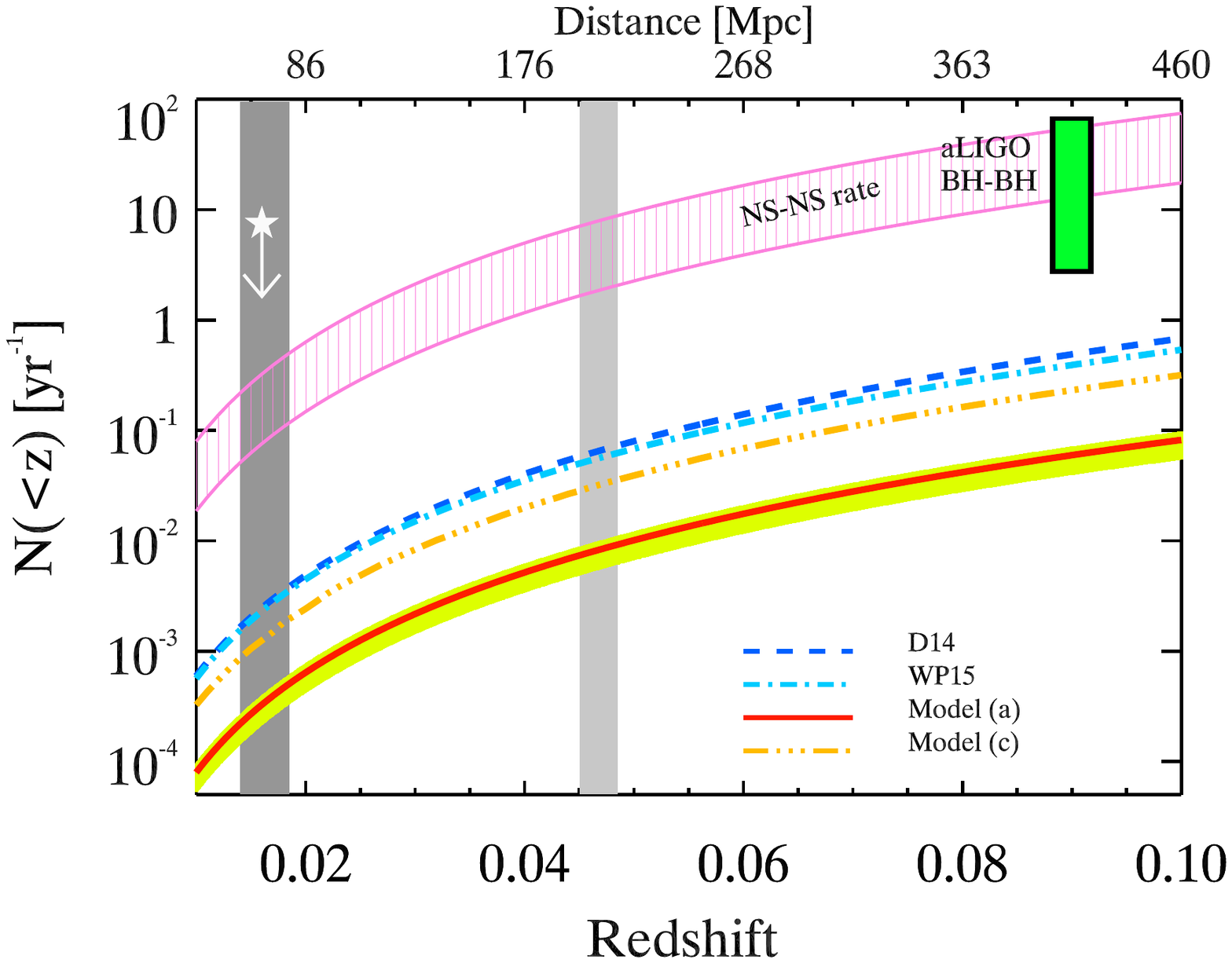} 
\vskip -7 truecm
\caption{Event rates within redshift $z$: solid red line and triple dot--dashed orange line represent the SGRB rates for case (a) and case (c) of this work, respectively. The yellow shaded region represents the 68\% confidence level on the rate (red line) of case (a). SGRB rates according to the models of D14 and WP15 are shown by the dashed blue and dot--dashed cyan lines, respectively. The rate of NS--NS mergers is shown by the hatched pink region where the lower (upper) boundary corresponds to the rate derived from population synthesis models (Galactic binaries) in \citealt{2015ApJ...806..263D} \citep{2015MNRAS.448..928K}. The vertical gray shaded regions show the present and design ranges of aLIGO for NS-NS mergers. The upper limit (white star) corresponds to the non--detection of NS--NS mergers in the first 48.6 days of the ``O1'' run of aLIGO. The green vertical bar is the  rate of binary BH mergers derived by \cite{2016arXiv160604856T} and shown here at the distance of GW150914 and GW151226.}
\label{fig:rate}
\end{figure*}

\section{Local SGRB rate}

The local rate of SGRBs is particularly important for the possible connection with gravitational wave events to be detected by the advanced interferometers (Advanced LIGO - \citealt{LIGO-Scientific-Collaboration:2015fr,Abbott:2016fk}; Advanced Virgo - \citealt{Acernese:2015zr}).


The first such detection, named GW150914, has been interpreted according to General Relativity as the space--time perturbation produced by the merger of two black holes (with masses $M_1\sim 29$ M$_{\odot}$ and $M_2\sim 36$ M$_{\odot}$) at a distance of $\sim$410 Mpc ($z = $0.09). 
The full analysis of the aLIGO first run cycle  revealed a second binary black hole merger event, GW151226 \citep{2016arXiv160604856T}. In this case the involved masses are smaller ($M_1\sim 14.2$ M$_{\odot}$ and $M_2\sim 7.5$ M$_{\odot}$) and the associated distance is only slightly larger ($\sim$440 Mpc)\footnote{A third event, LVT151012, was reported in \cite{2016arXiv160604856T} but with a small associated significance implying a probability of being of astrophysical origin  of $\sim$87\%.}.

GW150914 represents a challenge for the theory of formation and evolution of stellar origin BHs \citep{GW150914_astrophysical_implications,Belczynski:2016uq,Spera2016} being the most massive stellar-mass black hole observed so far. 
The masses of GW151226 are close to the ones observed in galactic X-ray binaries \citep{Ozel:2010db}. Both sources are an exquisite direct probe of General Relativity in the strong field dynamical sector \citep{GW150914_test_GR}. 

Considering the detections resulting from the analysis of the ``O1'' aLIGO interferometers, the rate of BH-BH merger is 9--240 Gpc$^{-3}$ yr$^{-1}$, assuming different BH mass distributions \citep{2016arXiv160604856T}. 
For the sake of comparison, in Fig. \ref{fig:rate} we show this range of rates (vertical green bar) in yr$^{-1}$ computed at the distance of GW150914.


However, the best is yet to come in the field of GW. Indeed, while no electromagnetic counterpart has been associated  either to GW150914 \citep[][but see \citealt{Connaughton:2016uq,2016ApJ...821L..18P,2016arXiv160205050Y,2016arXiv160204542Z,2016arXiv160205529M,2016arXiv160207352L}]{2016MNRAS.tmpL..45E,Troja:2016kx,2016arXiv160204156S,2016ApJ...820L..36S,2016arXiv160204198S,Annis:2016ys,Kasliwal:2016fr,Morokuma:2016zr,Ackermann:2016nx} and to GW151226 \citep{2016arXiv160604538C,2016arXiv160604795S,Adriani:2016qf,Evans:2016ul,Copperwheat:2016ve,Racusin:2016gf}, possible future detections of GW produced by compact binary mergers could lead to the first association of an electromagnetic with a gravitational signal \citep{Branchesi:2011vn,Metzger:2012fj}. In the case of NS--NS and NS--BH mergers, SGRBs are candidates to search for  
among other possible counterparts in the optical \citep{Metzger:2012fj}, X-ray \citep{Siegel:2016qy,Siegel:2016lq}, and radio bands \citep{Hotokezaka:2016uq}.

\begin{table}
\caption{Short GRB rates in $\rm{yr^{-1}}$ (68\% errors) within the volume corresponding to different distances: R = ``limiting distance for binary inspiral detection by aLIGO, averaged over sky location and binary inclination'', D = ``limiting distance for a face--on binary, averaged on sky location'', H = ``limiting distance (\textit{horizon}) for a face--on binary''. Limiting distances are obtained considering the aLIGO design sensitivity to NS--NS or NS--BH  inspirals (top and bottom portions of the table, respectively). \label{rate}}

\begin{center}
\renewcommand\arraystretch{1.3}
\begin{tabular}{cccc}
\hline
  & R & D & H \\
\hline\hline 

\it{NS--NS}  & $\le$200 Mpc                          &      $\le$300 Mpc                        &  $\le$450 Mpc			\\ 
Model (a) 		       & $0.007_{-0.003}^{+0.001}$    & 		$0.024_{-0.007}^{+0.004}$	&  $0.077_{-0.028}^{+0.014}$ \\
Model (c) 	        &  $0.028_{-0.010}^{+0.005}$    &		$0.095_{-0.034}^{+0.017}$	&  $0.299_{-0.108}^{+0.054}$ \\
\hline
\it{NS--BH}	&	$\le$410 Mpc		   	  &		      $\le$615 Mpc			&	$\le$927 Mpc			\\
Model (a)      		       & $0.060_{-0.022}^{+0.011}$    &		$0.20_{-0.07}^{+0.035}$      &  $0.572_{-0.206}^{+0.103}$ \\
Model (c)      		& $0.232_{-0.083}^{+0.042}$     &		$0.605_{-0.218}^{+0.109}$      &  $1.158_{-0.417}^{+0.208}$ \\
\hline
\end{tabular}
\end{center}
\end{table}

There is a considerable number of predictions for the rate of SGRBs within the horizon of GW detectors in the literature. The rather wide range of predictions, extending from 0.1 Gpc$^{-3}$ yr$^{-1}$ to $>200$ Gpc$^{-3}$ yr$^{-1}$ \citep[e.g.][]{2005A&A...435..421G,2006A&A...453..823G}, can be tested and further constrained by forthcoming GW-SGRB associations \citep{Coward:2014yq,Branchesi:2012rt}. If SGRBs have a jet, one must account for the collimation factor, i.e.\ multiply the rate by $f_b = \langle(1-\cos\theta_{\rm jet})^{-1}\rangle$, in order to compare such predictions with the compact binary merger rate. Once the luminosity function and rate of SGRBs is determined, the fraction of SGRBs above a limiting flux $P_{\rm min}$ within a given redshift $z$ is: 
\begin{equation}
N(<z)=\int_{0}^{z}dz\,C(z)\int_{L \ge L(P_{\rm min},z)}\phi(L)dL
\end{equation}
where $L(P_{\rm min},z)$ represents, at each redshift $z$, the minimum luminosity corresponding to the flux limit $P_{\rm lim}$ (e.g. of a particular GRB detector). 

Fig.~\ref{fig:rate} shows the rate of SGRBs within a given redshift $z$ (zoomed up to $z<0.1$). The different curves are obtained using the formation rate \psiz\ and luminosity function \lf\ by D14 and WP15 (shown by the dashed blue and dot-dashed cyan lines respectively) and the results of our case (a)  (red solid line) and case (c)  (triple dot--dashed orange line).  
 
These curves represent the population of SGRBs detectable in $\gamma$--rays by current flying instruments. At redshifts as low as those shown in Fig.~\ref{fig:rate}, even bursts populating the lowest end of the luminosity function can be observed above the flux limits of available GRB detectors (e.g. the \fe/GBM). The \psiz\ that we derive (see Fig.~\ref{fig:sfh_comparison}) rises, below the peak, in a way similar to those adopted in the literature (e.g. D14 and WP15). The lower rates predicted by our models with respect to those of D14 and WP15 are thus mainly due to our flatter \lf. 

The distance within which aLIGO should have been able to detect NS--NS mergers during ``O1'' was estimated to be $60$--$80\,\rm{Mpc}$, which corresponds to redshift $z\sim$0.014--0.0185 (dark grey shaded region in Fig.~\ref{fig:rate}) \citep{The-LIGO-Scientific-Collaboration:2016ys}.  
We use this distance to pose an upper limit on the NS--NS merger rate (star symbol and arrow in Fig.~\ref{fig:rate}), given the non detection of any such events in the 48.6 days of ``O1'' data \citep{The-LIGO-Scientific-Collaboration:2016ys}. 

If SGRBs have a jet, and if the jet is preferentially launched in the same direction as the orbital angular momentum, the inspiral of the progenitor binary could be detected up to a larger distance \citep[up to a factor $2.26$ larger, see][]{horizon_to_range_factor}, because the binary is more likely to be face--on. Let us define the following three typical distances: 
\begin{itemize}
 \item we indicate by R (\textit{range}) the limiting distance for the detection of a compact binary inspiral, averaged over all sky locations and over all binary inclinations with respect to the line of sight;
 \item we indicate by D (\textit{distance to face--on}) the limiting distance for the detection of a \textit{face--on} compact binary inspiral, averaged over all sky locations. ;
 \item we indicate by H (\textit{horizon}) the maximum limiting distance for the detection of a \textit{face--on} compact binary inspiral, i.e.\ the limiting distance at the best sky location.
\end{itemize}
Table~\ref{rate} shows R, D and H for both NS--NS binaries and BH--NS binaries, corresponding to the design sensitivity of Advanced LIGO, together with the expected rates of SGRBs (according to our models (a) and (c)) within the corresponding volumes.
The local rate of SGRBs predicted by our model (a) is $\rho_{0,a}=0.20^{+0.04}_{-0.07}$ yr$^{-1}$ Gpc$^{-3}$ and for model (c) $\rho_{0,c}=0.8^{+0.3}_{-0.15}$ yr$^{-1}$ Gpc$^{-3}$.
The distance R for NS--NS binary inspiral at design aLIGO sensitivity, which corresponds to 200 Mpc ($z\approx 0.045$), is shown by the vertical light gray shaded region in Fig.~\ref{fig:rate}. 

Fig.~\ref{fig:rate} also shows the predictions of population synthesis models for double NS merger \citep{2015ApJ...806..263D} or the estimates based on the Galactic population of NS \citep{2015MNRAS.448..928K} which bracket the pink dashed region in Fig.~\ref{fig:rate}. 

By comparing the SGRB models in Fig.\ref{fig:rate} with these putative progenitor curves, assuming that all NS--NS binary mergers yield a SGRB, we estimate the average jet opening angle of SGRBs as $\langle\theta_{\rm jet}\rangle\sim3^\circ -6^\circ$ in case (a) (solid red line in Fig.~\ref{fig:rate}). The local rates by D14 and WP15 instead lead to an average angle  $\langle\theta_{\rm jet}\rangle\sim7^\circ -14^\circ$. These estimates represent minimum values of the average jet opening angle, because they assume that all NS--NS binary mergers lead to a SGRB. We note that our range is  consistent with the very few SGRBs with an estimated jet opening angle: GRB 051221A ($\theta_{\rm jet}=7^\circ$, \citealt{2006ApJ...650..261S}), GRB 090426 ($\theta_{\rm jet}=5^\circ$, \citealt{2011A&A...531L...6N}), GRB 111020A ($\theta_{\rm jet}=3^\circ-8^\circ$, \citealt{2012ApJ...756..189F}), GRB 130603B ($\theta_{\rm jet}=4^\circ-8^\circ$, \citealt{2013ApJ...776...18F}) and GRB 140903A \citep{Troja:2016kx}. Similarly to the population of long GRBs \citep{2012MNRAS.420..483G}, the distribution of $\theta_{\rm jet}$ of SGRBs could be asymmetric with a tail extending towards large angles, i.e. consistently with the lower limits claimed by the absence of jet breaks in some SGRBs \citep{2014ARA&A..52...43B}.

\section{Conclusions}

We derived the luminosity function \lf, redshift distribution \psiz\ and local rate of SGRBs. Similarly to previous works present in the literature, we fitted the properties of a synthetic SGRB population, described by the parametric \lf\ and \psiz, to a set of observational constraints derived from the population of SGRBs detected by \fe\ and \sw. Any acceptable model of the SGRB population must reproduce their prompt emission properties and their redshift distributions. Our approach features a series of improvements with respect to previous works present in the literature: 
\begin{itemize}
\item (observer frame) constraints: we extend the classical set of observational constraints (peak flux and - for few events - redshift distribution) requiring that our model should reproduce the peak flux $P$, fluence $F$, peak energy \epo\ and duration $T$ distributions of 211 SGRBs with $P_{64}\geq 5\,\rm{ph\,s^{-1}\,cm^{-2}}$ as detected by the GBM instrument on board the \fe\ satellite. The uniform response of the GBM over a wide energy range (10 keV -- few MeV) ensures a good characterisation of the prompt emission spectral properties of the GRB population and, therefore, of the derived quantities, i.e. the peak flux and the fluence; 
\item (rest frame) constraints: we also require that our model reproduces the distributions of redshift, luminosity and energy of a small sample (11 events) of \sw\ SGRBs with $P_{64}\geq 3.5\,\rm{ph\,s^{-1}\,cm^{-2}}$ (selected by D14). This sample is 70\% complete in redshift and therefore it ensures a less pronounced impact of redshift--selection biases in the results; 
\item method: we parametrize \psiz\ as in Eq.~12 and derive the redshift distribution of SGRBs independently from their progenitor nature and their cosmic star formation history. Instead, the classical approach depends (i) on the assumption of a specific cosmic star formation history $\psi(z)$ and (ii) on the assumption of a delay time distribution $P(\tau)$;  
\item method: we derive our results assuming the existence of intrinsic \yone\ and \ama\ correlations in SGRBs (``case (a)''), similarly to what has been observed in the population of long GRBs. However, since evidence of the existence of such correlations in the population of SGRBs is still based on a limited number of bursts, we also explore the case of uncorrelated peak energy, luminosity and energy (``case (c)''). 
\end{itemize}
Our main results are: 
\begin{enumerate}
\item the luminosity function of SGRBs (case (a)), that we model with a broken power law, has a slope $\alpha_1 = 0.53^{+0.47}_{-0.14}$ (68\% confidence interval) below the break luminosity of $L_{\rm b} = 2.8^{+0.6}_{-1.89}\times 10^{52}$ erg s$^{-1}$ and falls steeply above the break with $\alpha_2 = 3.4^{+0.3}_{-1.7}$. This solution is almost independent from the specific assumption of the minimum luminosity of the \lf\ (case (b)). Moreover, it implies an average isotropic equivalent luminosity $\left\langle L \right\rangle \approx 1.5\times 10^{52}\,\rm{erg\,s^{-1}}$ (or $3\times 10^{52}\,\rm{erg\,s^{-1}}$ in case (c)), which is much larger than e.g.\ $\left\langle L \right\rangle \approx 3\times 10^{50}\,\rm{erg\,s^{-1}}$ from D14 or $\left\langle L \right\rangle \approx 4.5\times 10^{50}\,\rm{erg\,s^{-1}}$ from WP15;    
\item the redshift distribution of SGRBs \psiz\ peaks at $z\sim1.5$ and falls rapidly above the peak. This result is intermediate between those reported in the literature which assume either a constant large delay or a power law distribution favoring small delays. We find that our \psiz\ is consistent with the MD14 SFH retarded with a power law delay time distribution $\propto \tau^{-1}$; 
\item as a by-product we find that, if SGRBs feature intrinsic \yone\ and \ama\ correlations, they could be slightly steeper than those derived with the current small sample of short bursts  with redshift, e.g. \cite{Tsutsui:2013lr}, but still consistent within their 68\% confidence intervals;
\item if we assume that there are no correlations between \epo\ and \liso(\eiso) (case (c)), we find similarly that the \lf\ is flat at low luminosities and the formation rate peaks at slightly larger redshift ($z\sim 2$); 
  \item we estimate the rate of SGRBs as a function of $z$ within the explorable volume of advanced LIGO and Virgo for the detection of double NS mergers or NS--BH mergers. Assuming the design aLIGO sensitivity averaged over sky location and over binary orbital plane orientation with respect to the line of sight, NS--NS mergers can be detected up to 200 Mpc (410 Mpc for NS--BH mergers). This is usually referred to as the detection \textit{range} for these binaries. The rate of SGRBs within the corresponding volume is $\sim$7$\times10^{-3}$ yr$^{-1}$ (0.028 yr$^{-1}$ for NS--BH merger distance), assuming the existence of \yone\ and \ama\ correlations for the population of short bursts (model (a)).  Rates larger by a factor $\sim 4$ are obtained if no correlation is assumed (model (c)). If binaries producing observable SGRBs are preferentially face--on (which is the case if the GRB jet is preferentially aligned with the orbital angular momentum), then the actual explorable volume extends to a somewhat larger distance \citep[a factor of $\sim 1.5$ larger, see][]{schutz2011}, increasing the rates of coincident SGRB--GWs of about a factor of $3.4$ \citep{schutz2011};
\item we compare our SGRB rates with the rates of NS mergers derived from population synthesis models or from the statistics of Galactic binaries. This enables us to infer an average opening angle of the population of SGRBs of 3$^\circ$--6$^\circ$ (assuming that all SGRBs are produced by the NS--NS mergers) which is consistent with the few bursts with $\theta_{\rm jet}$ measured from the break of their afterglow light curve. 
\end{enumerate}

\noindent Our SGRB rate estimates might seem to compromise the perspective of a joint GW--SGRB observation in the near future. We note, though, that these rates refer to the prompt emission of SGRBs whose jets point towards the Earth. SGRBs not pointing at us can still be seen as ``orphan'' afterglows (i.e.\ afterglows without an associated prompt emission - see e.g. \citealt{Ghirlanda:2015fk,Rhoads1997} for the population of long GRBs) especially if the afterglow emission is poorly collimated or even isotropic \citep[e.g.][]{Ciolfi:2015kx}. The luminosity of the afterglow correlates with the jet kinetic energy, which is thought as proportional to the prompt luminosity. Point 1 above shows that the average luminosity in the prompt emission, as implied by our result, is higher by nearly two orders of magnitude than previous findings. This enhances the chance of observing an orphan afterglow in association to a GW event (e.g. \citealt{Metzger:2015fk}). Efforts should go in the direction of finding and identifying such orphan afterglows as counterparts of GW events.


\section*{Acknowledgments}
We acknowledge the financial support of the UnivEarthS Labex program at Sorbonne Paris Cit\'e (ANR-10-LABX-0023 and ANR-11-IDEX-0005-02) and the "programme PTV de l'Observatoire de Paris\' '' and GEPI for the financial support and kind hospitality during the implementation of part of this work. R.C. is supported by MIUR FIR Grant No. RBFR13QJYF. We acknowledge ASI grant I/004/11/1. We thank the referee for useful comments.

\bibliographystyle{aa}
\bibliography{journals,ghirlanda}

\end{document}